\documentstyle[preprint,prc,aps,epsfig]{revtex}
 \tighten
 \begin{document}
 \draft
\title{Neutron star properties in the quark-meson coupling model}
\author{S. Pal, M. Hanauske, I. Zakout, H. St\"ocker, and W. Greiner}
\address{Institut f\"ur Theoretische Physik, J.W. Goethe-Universit\"at,
60054 Frankfurt am Main, Germany}

\maketitle

\begin{abstract}

The effects of internal quark structure of baryons on the composition
and structure of neutron star matter with hyperons are investigated in 
the quark-meson coupling (QMC) model. The QMC model is based on mean-field 
description of nonoverlapping spherical bags bound by self-consistent 
exchange of scalar and vector mesons. The predictions of this model are
compared with quantum hadrodynamic (QHD) model calibrated to reproduce 
identical nuclear matter saturation properties. By employing a density
dependent bag constant through direct coupling to the scalar field, the 
QMC model is found to exhibit identical properties as QHD near saturation
density. Furthermore, this modified QMC model provides well-behaved and 
continuous solutions at high densities relevant to the core of neutron 
stars. Two additional strange mesons are introduced which couple
only to the strange quark in the QMC model and to the hyperons in the 
QHD model. The constitution and structure of stars with hyperons in the 
QMC and QHD models reveal interesting differences. This suggests the 
importance of quark structure effects in the baryons at high densities.
\end{abstract}



\section{Introduction}
 
Neutron stars are born in the aftermath of supernova explosions with 
interior temperatures $T\stackrel{>}{\sim} 10^{11}$ K, but cool 
rapidly in a few seconds by deleptonization \cite{Bur} to almost cold 
nuclear matter. Neutron star matter is charge neutral so that gravitational 
force can bound it against the relatively strong repulsive Coulomb
force, and is in $\beta$-equilibrium condition, i.e. in its lowest
energy state. Since the matter density in the core could exceed a few
times the normal nuclear matter density, neutron star matter provides
an interesting possibility to investigate the strong interaction effects
which are poorly understood at supernuclear density. In fact, the structure
of a neutron star is chiefly determined by the equation of state (EOS) of
the strongly interacting constituents.

There have been several attempts to determine the EOS for dense nucleon
matter which are based primarily on nonrelativistic potential models and
relativistic field theoretical models. The former approach comprises 
a Hamiltonian with a two-nucleon potential fitted to nucleon-nucleon
scattering data and the properties of deuteron. The quantum many-body
problem is traditionally handled either by a selective summation of
diagrams in perturbation theory (the Brueckner-Bethe-Goldstone approach)
or by using a variational method with correlation operators \cite{Fri,Wir}.
However, an important shortcoming of many potential models are that they
are well suited at low densities only; the EOS becomes acausal i.e., the
speed of sound exceeds that of light at densities relevant for maximum
mass neutron stars. Moreover, these models lead to symmetry energies that
drastically decreases beyond about three times the saturation density which
is a serious deficiency for highly asymmetric systems like neutron stars.
The alternative approach to the nuclear many body problem involves the
formulation of effective relativistic field theory within the framework of
quantum hadrodynamics (QHD) \cite{Ser,Chi} where the appropriate degrees 
of freedom are
the baryons interacting through the exchange of isoscalar scalar and vector
mesons ($\sigma, \omega$) and the isovector vector $\rho$ meson which
provides the driving force to the isospin symmetry. The theory is an effective
one since the coupling constants are determined by the saturation properties
of nuclear matter. The equations of motion for the baryons and mesons are
solved self-consistently in the mean field approximation. There exists a
large body of calculations based on QHD which were found to provide a 
realistic description of bulk properties of finite nuclei and nuclear 
matter \cite{Ser,Chi,Rei,Ser97}. A large number of single-particle 
properties of finite nuclei are also well accounted including the charge 
distributions, spin-orbit interactions etc. \cite{Rei,Ser97,Fur87,Bod,Gam}.
The central to the success of the theory is that the small binding energy
in a nucleus arises from the cancelation between large Lorentz scalar
and vector potentials, each of which is approximately several hundred
MeV at saturation density. The potentials being comparable to the nucleon
mass, it indicates the importance of relativistic effects even at normal
densities. With increasing density the Fermi momentum of the baryons 
increases further which points to stronger relativistic effects. Therefore
QHD appears to be more appropriate than nonrelativistic models for neutron 
star calculations. Moreover, the Lorentz covariance in this theory is retained
from the outset. Consequently, the EOS automatically respect the causality
limit. The relativistic mean field models with baryons and mesons as point
particles, and the coupling constants determined from the saturation 
properties of nuclear matter, has been extrapolated into the high density 
regime to investigate the neutron star properties 
\cite{Gle,Kno,Sch96,Pra97,Ban}.

At high densities in the core of a neutron star the relevant degrees of
freedom that may be crucial to its structure and composition are the 
effect of quarks confined within the hadrons. Small but interesting corrections
to the standard hadronic picture has been already observed such as the
EMC effect which reveals the medium modification of the internal structure
of the nucleon \cite{Arn}. By treating the nucleons as structureless particles,
the QHD model completely misses the important effect of the constituent quarks
especially at large densities. It is therefore instructive to study the
relevance of the quark structure of the nucleon at supernuclear density
regime as in the neutron stars. Quantum chromodynamics (QCD) which governs
the underlying strong interactions of quarks and gluons, although, believed
to be essentially the fundamental theory at the nuclear and subnuclear scale
is however intractable due to the nonperturbative features of QCD. It is
therefore not a theory from which one can derive practical results for the
equation of state. The most fruitful way to deal this situation would 
be to approach the problem from both sides, i.e. to develop the QHD and QCD
motivated models and to compare their predictions.
 
The spontaneous symmetry breaking and chiral symmetry restoration criteria
have been utilized to model effective Lagrangians for the low energy 
strong interaction, as for example the Nambu$-$Jona-Lasinio model \cite{Nam}.
However, these models have never been applied successfully for the description
of the saturation properties of nuclear matter. On the other hand, the
quark-meson coupling (QMC) model proposed by Guichon \cite{Gui88} provides
a simple and attractive framework to investigate the direct quark effects
in nuclei. The model describes nuclear matter as nonoverlapping, spherical
and static MIT bags in which the quarks interact through a self-consistent
exchange of structureless scalar $\sigma$ and vector $\omega$ mesons in the
mean field approximation. This simple QMC model has been subsequently refined 
by including nucleon Fermi motion and center of mass corrections to the bag
energy \cite{Fle} and applied with reasonable success to various problems 
of nuclear matter \cite{Sai94,Tho,Sai,Son,Jin,Jin2} and finite nuclei 
\cite{Gui96,Blu,Mul98}. Recently, the model has also been used to investigate 
the properties of $\Lambda$, $\Sigma$ and $\Xi$ hypernuclei \cite{Tsu}.

Although it provides a simple and interesting framework to describe 
the basic features of the nuclear systems in term of quark degrees freedom,
the QMC model has two serious shortcomings. Firstly, it predicts much
smaller scalar and vector potentials for the nucleon than that obtained
in the well-established QHD model. This implies a much smaller nucleon 
spin-orbit potential (sum of scalar and vector potentials) and hence weaker
spin-orbit splittings in finite nuclei. Secondly, it is not apparent how 
far the EOS in the QMC model can be extrapolated at high densities in the 
neutron star interior because the assumption of nonoverlapping bags may 
break down. It was recently pointed out \cite{Jin,Jin2} that the small 
values of the nucleon potentials stem from the assumption that
the bag constant is fixed at its free space value. By introducing a density
dependent bag constant (by considering it to be proportional to higher powers
of the scalar field $\sigma$, for example) it was demonstrated \cite{Jin2}
that large scalar and vector potentials are obtained. Evidently, a drop
in the bag value with increasing density implies a decrease in the effective
nucleon mass in a self-consistent manner. Furthermore, it was shown that the 
quark substructure is entirely contained in the scalar potential \cite{Sai94}.
Therefore, the density dependent bag constant through $\sigma$ field
represents, in a way, self-interaction in the scalar field in this model.
This feature of nonlinear interaction in sigma field, obtained by $\sigma$
field dependent bag constant may also remedy the other shortcoming of 
the overlapping bags at high density matter. One possible way to verify that 
the QMC model may be applied at supernuclear densities is by exploring 
observables as a function of density which shall not deviate
from its continuous behavior at high densities. In other words, the QMC 
theory which provides a reasonable behavior of the equation of state etc.
in the vicinity of saturation, should also be well-behaved when 
extrapolated to supernuclear density regime. Of course, the QMC model
at densities near nuclear matter value should reproduce the results based
on the more established QHD model. Any deviation (not discontinuity) in 
the QMC results from that of the QHD at high densities may then be regarded as 
the effects of the quark structure beyond the structureless hadronic picture.

In this paper we shall show that by employing a bag constant coupled to the
$\sigma$ field, the continuity of the equation of state etc. indeed persists
in the QMC model when extrapolated to high densities relevant to the neutron
star core. This will allow us to investigate the structure and composition of 
a neutron star in the QMC model and compare its predictions with that in the 
QHD model. Since the QMC model with bag constant as function of $\sigma$ field
represents nonlinear scalar self-interactions, for comparison we employ a 
QHD model which contains cubic and quartic scalar self-interactions 
\cite{Bog}; the latter model is calibrated to reproduce the same saturation 
properties as the QMC model.

At high densities and hence large nucleon Fermi energies expected in the
cores of stars, weak interaction energetically favor the 
conversion of some nucleons at the Fermi surface to hyperons. The 
negatively charged hyperons can then also replace the leptons. These
effects cumulatively can decrease the pressure resulting in the softening
of the equation of state. Within the QMC framework, only hypernuclear
matter and hypernuclei \cite{Sai,Tsu} (without charge neutrality and 
$\beta$-equilibrium conditions) have been investigated where the bag 
constant is fixed to its free space value. Moreover, it was assumed 
\cite{Sai,Tsu} that the nonstrange $\sigma$ and $\omega$ mesons couple only 
to the $u$ and $d$ quarks; the $s$ quark is unaffected in the medium and set 
to its constant bare mass value. We investigate neutron star properties with 
an improved Lagrangian by incorporating an additional pair of hidden strange 
meson fields ($\sigma^*$, $\phi$) which couple only to the $s$ quark in the 
QMC model and only to the hyperons in the QHD model. The standard mean field 
model was found to be inadequate to describe the strongly attractive 
hyperon-hyperon interaction observed in double $\Lambda$ hypernuclei 
\cite{Sch94}. The hyperon-hyperon interaction is expected to be important 
for hyperon rich matter \cite{Gle} present in the cores of neutron stars.

The paper is organized as follows. In section II, we introduce the 
relativistic mean field models, the QMC and the QHD, both extended to 
incorporate the two additional (hidden) strange mesons. The relevant equations
for neutron star matter with hyperons are summarized in these models.
In section III, we make a systematic comparison of the QMC and the QHD 
results in the entire density range relevant to neutron stars. In section IV,
we use the model to study the constitution and structure of a star. 
Section V is devoted to summary and conclusions.

\section{The Relativistic Mean Field Models}

\subsection{The Extended Quark-Meson Coupling Model}

In this section we present a brief introduction to the quark meson
coupling model for baryonic matter and its extension to include the hidden
strange mesons which couple explicitly only to the $s$ quark in a hyperon bag.
In the QMC model, a baryon in nuclear medium is assumed to be a static 
spherical MIT bag in which the quarks are coupled to the meson fields in the 
relativistic mean field (RMF) approximation. In QMC model calculations for 
hadronic matter \cite{Sai,Tsu}, the meson fields considered are isoscalar 
scalar $\sigma$ and vector $\omega$ mesons and the isovector vector
$\vec{\rho}^\mu$ meson. However, they being nonstrange are
allowed to couple only to the $u$ and $d$ quarks within a baryon bag, while
the $s$ quark is unaffected. It is expected that with increasing density,
the $s$ quark mass should also be modified. To implement this situation, we 
incorporate two additional mesons, the scalar meson $f_0(m=975$ MeV)
(denoted as $\sigma^*$ hereafter) and the vector meson $\phi(m=1020$ MeV)
with their masses given in the parenthesis. These strange mesons couple only 
to the $s$ quark in a hyperon bag. This extended QMC model has an additional 
advantage that it accounts for the strongly attractive $\Lambda\Lambda$ 
interaction observed in hypernuclei which cannot be reproduced by 
($\sigma$, $\omega$, $\rho$) mesons only \cite{Sch94}.

For a uniform static matter within the RMF, let the mean fields be denoted
by $\sigma$, $\sigma^*$ for the scalar mesons, and $\omega_0$, $\phi_0$
and $\rho_{03}$ for the timelike and the isospin 3-component of the
vector and the vector-isovector mesons. The Dirac equation for a quark
field $\psi_q$ in a bag is then given by
\begin{equation}
\left[ i\gamma\cdot\partial - \left(m_q - g^q_\sigma \sigma
- g^q_{\sigma^*} \sigma^* \right) - \gamma^0\left(g^q_\omega \omega_0
 + g^q_\phi \phi_0 + \frac{1}{2} g^q_\rho \tau_z \rho_{03} \right)
\right] \psi_q = 0 ~,
\end{equation}
where $g^q_\sigma$, $g^q_{\sigma^*}$, $g^q_\omega$, $g^q_\phi$, $g^q_\rho$ 
are the quark-meson coupling constants and $m_q$ the bare mass of the
quark $q\equiv (u,d,s)$. The normalized ground state for a quark in a bag
is given by
\begin{equation}
\psi_q({\bf r}, t) = {\cal N}_q \exp \left(-i\varepsilon_q t/R_B \right)
\left( \matrix{ j_0\left(x_q r/R_B\right)\cr
i\beta_q \vec{\sigma} \cdot \hat r j_1\left(x_q r/R_B\right) }
\right) \frac{\chi_q}{\sqrt{4\pi}} ~,
\end{equation}
where
\begin{equation}
\varepsilon_q = \Omega_q + R_B\left( g^q_\omega \omega_0 + g^q_\phi \phi_0 
+ \frac{1}{2} g^q_\rho \tau_z \rho_{03} \right) ~; ~~~
\beta_q = \sqrt{\frac{\Omega_q - R_B m_q^*}{\Omega_q + R_B m_q^*}} ~.
\end{equation}
The normalization factor is given by
\begin{equation}
{\cal N}_q^{-2} = 2R_B^3 j_0^2(x_q)\left[\Omega_q(\Omega_q-1) 
+ R_B m_q^*/2 \right] \Big/ x_q^2 ~,
\end{equation}
with $\Omega_q = \sqrt{x_q^2 + (R_Bm_q^*)^2}$ the kinetic energy of the
quark $q$ and $R_B$ is the radius of a baryon $B$, and $\chi_q$ the quark
spinor. The effective mass of a quark is given by
\begin{equation}
 m_q^* = m_q - g^q_\sigma \sigma - g^q_{\sigma^*} \sigma^* ~.
\end{equation}
The linear boundary condition, $j_0(x_q) = \beta_q j_1(x_q)$, at the
bag surface determines the eigenvalue $x_q$.

The hyperon couplings are not relevant to the ground state properties
of nuclear matter, but information about them can be available
from levels in $\Lambda$-hypernuclei \cite{Chr}. Experimental data of
$\Sigma$-hypernuclei are scarce and ambiguous because of the strong 
$\Sigma N \to \Lambda N$ decay, while only few events in emulsion 
experiments with $K^-$ beams have been attributed to the formation of
$\Xi^-$ hypernuclei. Therefore considerable uncertainty in the 
hyperon-nucleon interaction exists even at the normal nuclear density, 
their interactions at high densities are more ambiguous. In view of the
uncertainties in the hyperon couplings, for simplicity, we
employ in this paper the SU(6) symmetry based on the light ($u$, $d$) quark 
counting rule for both the scalar ($\sigma$, $\sigma^*$) and vector 
($\omega$, $\phi$) coupling constants to the hyperons. The coupling 
constants are thus related by
\begin{eqnarray}
& & \frac{1}{3}g_{\sigma N} = \frac{1}{2}g_{\sigma\Lambda}
= \frac{1}{2}g_{\sigma\Sigma} = g_{\sigma\Xi} \equiv g^q_\sigma ~,\nonumber \\
& & \frac{1}{3}g_{\omega N} = \frac{1}{2}g_{\omega\Lambda}
= \frac{1}{2}g_{\omega\Sigma} = g_{\omega\Xi} \equiv g^q_\omega ~, \nonumber \\
& & g_{\rho N} = \frac{1}{2}g_{\rho\Sigma} = g_{\rho\Xi} \equiv g^q_\rho ~, 
~~~ g_{\rho\Lambda} = 0 ~.
\end{eqnarray}
The couplings to the strange mesons are
\begin{eqnarray}
& & 2g_{\sigma^*\Lambda} = 2g_{\sigma^*\Sigma}
= g_{\sigma^*\Xi} = \frac{2\sqrt 2}{3}g_{\sigma N} 
\equiv 2\sqrt{2}g^q_{\sigma} ~, ~~~~ g_{\sigma^* N} = 0 \nonumber \\
& & 2g_{\phi\Lambda} = 2g_{\phi\Sigma}
= g_{\phi\Xi} = \frac{2\sqrt 2}{3}g_{\omega N} 
\equiv 2\sqrt{2}g^q_{\omega} ~, ~~~~ g_{\phi N} = 0 ~. 
\end{eqnarray}
Note that in a baryon the $u$ and $d$ quarks are not coupled to the strange 
($\sigma^*$, $\phi$) mesons i.e., $g^{u,d}_{\sigma^*} = g^{u,d}_{\phi} = 0$, 
while the $s$ quark is unaffected by the ($\sigma$, $\omega$) mesons 
i.e., $g^s_\sigma = g^s_\omega = 0$. Furthermore, from Eqs. (6) and (7)
it is evident that the coupling constants of the $s$ quark to 
($\sigma^*$, $\phi$) may be obtained from the coupling constants of the 
($u$, $d$) quarks to ($\sigma$, $\omega$) mesons by the relation
$g^s_{\sigma^*} = {\sqrt 2} g^{u,d}_{\sigma}$ and
$g^s_{\phi} = {\sqrt 2} g^{u,d}_{\omega}$. The energy of a baryon bag 
consisting of three ground state quarks is then given by 
\begin{equation}
E_B^{\rm bag} = \frac{\sum_q n_q\Omega_q - z_B}{R_B} 
+ \frac{4}{3}\pi R_B^3 B_B ~,
\end{equation}
where $n_q$ is the number of quarks of type $q$, $z_B$ accounts for the
zero-point motion, and $B_B$ is the bag constant for the baryon species $B$.
After the corrections of spurious center of mass motion, the effective
mass of a baryon is given by \cite{Fle,Sai94}
\begin{equation}
m_B^* = \sqrt{ (E^{\rm bag}_B)^2 - \sum_q n_q(x_q/R_B)^2 } ~.
\end{equation}
For fixed meson fields, the bag radius $R_B$ is determined by the
equilibrium condition of the baryon bag in the medium 
$\partial m_B^* / \partial R_B = 0$. For a given value of bag constant
$B_B = B_0$ in free space, the parameter $z_B$ and the bag radius $R_B$ of
the baryons may be obtained by reproducing the physical masses of the baryons
i.e. Eq. (9) in free space. In the present calculation, the current
quark masses considered are $m_u = m_d = 0$ and $m_s = 150$ MeV. For
our choice of free space bag constant  $B_0^{1/4} = 188.1$ MeV, the
values of $z_B$ and $R_B$ are collected in Table I. It is seen that for
the fixed bag value, the equilibrium condition in free space results 
in an increase of the bag radius and a decrease of the zero-point 
motion for the heavier species.

In the original version of the QMC model \cite{Gui88} the bag constant
was held fixed at its free space value $B = B_0$. The bag constant is a 
nonuniversal quantity associated with the QCD trace anomaly. When a baryonic
bag is immersed in matter, it is expected to decrease from its free space
value as argued in Ref. \cite{Ada}. At present, however, no reliable 
information on the medium dependence of $B_B$ is available on the level of
QCD calculations. Effective models, as for example the Nambu$-$Jona-Lasinio
(NJL) model \cite{Nam} which approximate low energy QCD are constructed based 
on symmetries and symmetry breaking patterns of QCD, in particular, the chiral
symmetry breaking. The concept of bag constant arises naturally in these
models where its value decreases when the density of the nuclear environment
is increased \cite{Asa}. To reflect this physics in the QMC model, the density
dependence of the bag constant for a nucleon was proposed by Jin and Jennings 
\cite{Jin2} by considering its direct coupling to the scalar mean field, i.e.,
\begin{equation}
B_N/B_0 = \exp\left[-4g_\sigma^B \sigma \Big/m_N \right] ~,
\end{equation}
where $g_\sigma^B$ is a real parameter; in this paper we shall adopt this 
form of exponential dependence. This direct coupling model is inspired by 
NJL type nontopological soliton model for the nucleon \cite{Alk}
where a scalar soliton field is responsible for the confinement of the
quarks to form a nucleon. When the nucleon soliton is inserted into the
nuclear environment, the scalar soliton field will interact with the
scalar mean field \cite{Bane}. In fact, a similar approach is used to construct
in QHD the nonlinear mean field models, where the unknown density dependence
of the nuclear energy functional is parametrized by nonlinear meson-meson
interactions \cite{Fur96,Mul96}. The direct coupling of the bag constant
to the scalar mean field $\sigma$ in nucleonic medium needs to be extended
for hyperons where additional scalar field $\sigma^*$ is employed. To this end,
we employ consistently the SU(6) symmetries for the scalar couplings of 
Eqs. (6) and (7). Specifically, the bag constant $B_B$ of a baryon is
directly coupled to $\sigma$ and $\sigma^*$ fields through the relation
\begin{equation}
B_B/B_0 = \exp\left[-4g_\sigma^{'B} \left( \sum_{q=u,d} n_q \sigma
+ \Big(3 - \sum_{q=u,d} n_q \Big) \sqrt{2} \sigma^* \right)
\Bigg/m_B \right] ~,
\end{equation}
where $m_B$ is the bare mass of the baryon $B$, and for nucleonic bags 
$\sum n_q=3$. This modeling has only one real positive parameter
$g_\sigma^{'B}$ which for nucleon is related to $g_\sigma^B$ of Eq. (10)
by $g_\sigma^{'B} = g_\sigma^B/3$. We have refrained from using an additional
parameter for coupling to the $\sigma^*$ and have rather used the SU(6) 
symmetries because we believe that a reliable extrapolation to high densities
should be based on a model having as few adjustable parameters as possible so 
that the model having been fitted to saturation properties of nuclear
matter can be tested for its predictive power under conditions not included
in the determination of the parameters. It may be also noted that in the 
parametrization (11), the use of free baryonic mass $m_B$ is essential. By
considering only ($\sigma$, $\omega$, $\rho$) mesons in the model, we have
found that this choice of $B_B/B_0$ results at nuclear matter density the 
scaling relation $\delta m^*_{\Lambda,\Sigma} / \delta m^*_N \approx 2/3$
and $\delta m^*_\Xi / \delta m^*_N \approx 1/3$, where
$\delta m^*_B = m_B - m_B^*$ for the baryon $B$. If we now define the field
dependent $\sigma-B$ coupling constant $g_{\sigma B}(\sigma)$ by
\begin{equation}
g_{\sigma B}(\sigma)\sigma = m_B - m_B^*(\sigma) ~,
\end{equation}
the same scaling relation is obtained at the nuclear matter density i.e.,
$\delta m^*_B/\delta m^*_N = g_{\sigma B}(\sigma)/g_{\sigma N}(\sigma)$
are $2/3$ and $1/3$ for ($\Lambda$, $\Sigma$) and $\Xi$, respectively.
This implies that the parametrization (11) in the nonstrange sector is
consistent with the SU(6) symmetry employed in determining the $\sigma-B$
coupling constants. In this model the baryon effective mass $m_B^*$ and 
the bag constant are determined self-consistently by combining Eqs. 
(8), (9) and (11). In principle, the parameter $z_B$ may also be modified
in baryonic medium. However, unlike the bag constant it is not clear
how $z_B$ changes with density as it is not directly related to chiral
symmetry. In this paper we assume $z_B$ remains constant at its free
space value $z_B = z_0$ as given in Table I.

The total Lagrangian density of a neutron star matter for the full
baryon octet in the QMC model within RMF approximation can be written as
\begin{eqnarray}
{\cal L}_{\rm QMC} &=& \sum_B {\overline \psi}_B \left[ i\gamma\cdot\partial
- m_B^*(\sigma,\sigma^*) - \gamma^0 \left( g_{\omega B}\omega_0
+ g_{\phi B}\phi_0 + \frac{1}{2} g_{\rho B}\tau_z\rho_{03} \right) 
\right] \psi_B  \nonumber \\
&+& \frac{1}{2}\left(m_\sigma^2\sigma^2 + m_{\sigma^*}^2\sigma^{* 2} 
+ m_\omega^2\omega_0^2 + m_\phi^2\phi_0^2 + m_\rho^2\rho_{03}^2 \right) 
+ \sum_l {\overline \psi}_l \left(i\gamma\cdot\partial - m_l\right) \psi_l ~.
\end{eqnarray}
Here the sum on B is over all the charge states of the baryon octet
($p, n, \Lambda, \Sigma^+, \Sigma^0, \Sigma^-, \Xi^0, \Xi^-$) coupled to the
$\sigma, \omega, \rho, \sigma^*, \phi$ mesons. The sum on $l$ is over
the free electrons and muons ($e^-$ and $\mu^-$) in the star. The baryon 
effective mass of Eq. (9) may be expressed in terms of the field
dependent $\sigma-B$ and $\sigma^*-B$ coupling strengths
$g_{\sigma B}(\sigma)$ and $g_{\sigma^* B}(\sigma^*)$ as
\begin{equation}
m_B^*(\sigma,\sigma^*) = m_B -  g_{\sigma B}(\sigma)\sigma 
-  g_{\sigma^* B}(\sigma^*)\sigma^*  ~. 
\end{equation}
The dependences of the coupling strengths on the applied scalar fields
must be calculated self-consistently within the quark model.

For the QMC model, the equations of motion for the meson fields in
uniform static matter are given by
\begin{equation}
m_\sigma^2\sigma = \sum_B g_{\sigma B} C_B(\sigma) \frac{2J_B + 1}{2\pi^2}
\int_0^{k_B} \frac{m_B^*(\sigma,\sigma^*)}
{\left[k^2 + m_B^{* 2}(\sigma,\sigma^*)\right]^{1/2}} \: k^2 \ dk ~,
\end{equation}
\begin{equation}
m_{\sigma^*}^2\sigma^* = \sum_B g_{\sigma^* B} C_B(\sigma^*) 
\frac{2J_B + 1}{2\pi^2} \int_0^{k_B} \frac{m_B^*(\sigma,\sigma^*)}
{\left[k^2 + m_B^{* 2}(\sigma,\sigma^*)\right]^{1/2}} \: k^2 \ dk ~,
\end{equation}
\begin{equation}
m_\omega^2\omega_0 = \sum_B g_{\omega B} \left(2J_B + 1\right)
k_B^3 \big/ (6\pi^2) ~,
\end{equation}
\begin{equation}
m_\phi^2\phi_0 = \sum_B g_{\phi B} \left(2J_B + 1\right)
k_B^3 \big/ (6\pi^2) ~,
\end{equation}
\begin{equation}
m_\rho^2\rho_{03} = \sum_B g_{\rho B} I_{3B} \left(2J_B + 1\right)
k_B^3 \big/ (6\pi^2) ~.
\end{equation}
In the above equations $J_B$ and $I_{3B}$ are the spin and the isospin 
projection and $k_B$ is the Fermi momentum of the baryon species $B$. On 
the right hand sides of Eqs. (15) and (16), a new characteristic feature 
of QMC beyond QHD appears through the factors $C_B(\sigma)$ and 
$C_B(\sigma^*)$ where
\begin{eqnarray}
g_{\sigma B}C_B(\sigma) = - \frac{\partial m_B^*(\sigma,\sigma^*)}
{\partial \sigma} &=& \sum_{q=u,d} n_q \Bigg[ g^q_\sigma \frac{E_B^{\rm bag}}
{m_B^*(\sigma,\sigma^*)} \left\{ \left(1 - \frac{\Omega_q}{E_B^{\rm bag}R_B}
\right) S_B(\sigma) + \frac{m_q^*}{E_B^{\rm bag}} \right\} \nonumber\\
&+& g_\sigma^{' B} \frac{E_B^{\rm bag}}{m_B^*(\sigma,\sigma^*)} 
\frac{16}{3} \pi R_B^3 \frac{B_B}{m_B} \Bigg] ~,
\end{eqnarray}
\begin{eqnarray}
g_{\sigma^* B}C_B(\sigma^*) = - \frac{\partial m_B^*(\sigma,\sigma^*)}
{\partial \sigma^*} &=& n_s g^q_{\sigma^*} \frac{E_B^{\rm bag}}
{m_B^*(\sigma,\sigma^*)} \left\{ \left(1 - \frac{\Omega_s}{E_B^{\rm bag}R_B}
\right) S_B(\sigma^*) + \frac{m_s^*}{E_B^{\rm bag}} \right\} \nonumber\\
&+& n_s g_\sigma^{' B} \frac{E_B^{\rm bag}}{m_B^*(\sigma,\sigma^*)} 
{\sqrt 2} \frac{16}{3} \pi R_B^3 \frac{B_B}{m_B} ~,
\end{eqnarray}
where $n_s$ ($=3-\sum_{q=u,d} n_q$) is the number of $s$ quark in the
baryon. The quark scalar densities in the bag are
\begin{equation}
S_B(\sigma) = \int_{k_B} d{\bf r} \ {\overline \psi}_q \psi_q 
= \frac{\Omega_q/2 + R_Bm^*_q(\Omega_q - 1)}
{\Omega_q(\Omega_q -1) + R_Bm_q^*/2} ~; ~~~~ q \equiv (u,d) ~,
\end{equation}
and a similar expression for $S_B(\sigma^*)$ for the contribution
from the medium modification of $s$ quark in the field $\sigma^*$.
The medium dependence of the scalar densities on the bag radius was found
to be rather insensitive. The last terms in Eqs. (20) and (21) originate
from the density dependent bag constants through direct coupling to the scalar
fields. It is evident from Eqs. (11), (15) and (16) that an exponential
dependence of the bag constant on the fields $\sigma$ and $\sigma^*$
introduce a nonlinear self-interaction in these fields. 
Moreover, the decrease of $C_B$ with increasing density provides a new
source of attraction and thereby constitutes a new saturation mechanism
which is different from QHD.

For stars in which the strongly interacting particles are baryons, the
composition is determined by the requirements of charge neutrality
and $\beta$-equilibrium conditions under the weak processes 
$B_1 \to B_2 + l + {\overline \nu}_l$ and $B_2 + l \to B_1 + \nu_l$.
Under the conditions that the neutrinos have left the system, the
charge neutrality condition gives
\begin{equation}
q_{\rm tot} = \sum_B q_B (2J_B + 1) k_B^3 \big/ (6\pi^2)
+ \sum_{l=e,\mu} q_l k_l^3 \big/ (3\pi^2)  = 0 ~,
\end{equation}
where $q_i$ corresponds to the electric charge of species $i$. Since
the time scale of a star is effectively infinite compared to the weak
interaction time scale, weak interaction violate strangeness conservation.
The strangeness quantum number is therefore not conserved 
in a star and the net strangeness is determined by the condition of 
$\beta$-equilibrium which for baryon $B$ is then given by
$\mu_B = b_B\mu_n - q_B\mu_e$, where $\mu_B$ is the chemical potential
of baryon $B$ and $b_B$ its baryon number. Thus the chemical potential of any
baryon can be obtained from the two independent chemical potentials $\mu_n$
and $\mu_e$ of neutron and electron. The Fermi momentum of the baryons
can be obtained from the solution of the equation $\varepsilon_B(k_B)=\mu_B$,
where the energy eigenvalues of the Dirac equation for the baryons are
\begin{equation}
\varepsilon_B(k) = \sqrt{k^2 + m_B^{* 2}(\sigma,\sigma^*)}
+ g_{\omega B}\omega_0 +  g_{\phi B}\phi_0 + g_{\rho B}I_{3B}\rho_{03} ~. 
\end{equation}
The lepton Fermi momenta are the positive real solutions of 
$(k_e^2 + m_e^2)^{1/2} =  \mu_e$ and
$(k_\mu^2 + m_\mu^2)^{1/2} = \mu_\mu = \mu_e$. The equilibrium composition 
of the star is obtained by solving the set of Eqs. (15)-(19) in conjunction 
with the charge neutrality condition (23) at a given total baryonic density
$n_B = \sum_B b_B (2J_B + 1) k_B^3/(6\pi^2)$; the baryon effective masses are 
obtained self-consistently in the bag model. The total energy density and
pressure including the leptons can be obtained from the grand canonical 
potential to be 
\begin{eqnarray}
\varepsilon &=& \frac{1}{2}m_\sigma^2 \sigma^2 
+ \frac{1}{2}m_{\sigma^*}^2 \sigma^{* 2} + \frac{1}{2}m_\omega^2 \omega^2_0
+ \frac{1}{2} m_\phi^2 \phi^2_0 + \frac{1}{2} m_\rho^2 \rho^2_{03} \nonumber\\
&+& \sum_B \frac{2J_B +1}{2\pi^2} \int_0^{k_B}
\left[k^2 + m_B^{* 2}(\sigma,\sigma^*)\right]^{1/2} k^2 \ dk
+ \sum_l \frac{1}{\pi^2} \int_0^{k_l} \left[k^2 + m_l^2\right]^{1/2} k^2 \ dk ~,
\end{eqnarray}
\begin{eqnarray}
P &=& - \frac{1}{2}m_\sigma^2 \sigma^2 
- \frac{1}{2}m_{\sigma^*}^2 \sigma^{* 2} + \frac{1}{2}m_\omega^2 \omega^2_0
+ \frac{1}{2} m_\phi^2 \phi^2_0 + \frac{1}{2} m_\rho^2 \rho^2_{03} \nonumber\\
&+& \frac{1}{3} \sum_B \frac{2J_B +1}{2\pi^2} \int_0^{k_B}
\frac{k^4 \ dk}{\left[k^2 + m_B^{* 2}(\sigma,\sigma^*)\right]^{1/2}}
+ \frac{1}{3} \sum_l \frac{1}{\pi^2} \int_0^{k_l} \frac{k^4 \ dk}
{\left[k^2 + m_l^2\right]^{1/2}} ~.
\end{eqnarray}

To obtain the coupling constants and the parameters in the QMC model, we recall
that for a fixed $B_0^{1/4} = 188.1$ MeV, the $z_0$ values have been adjusted
to reproduce the baryon masses in free space and these are listed in Table I.
For a given value of $g^q_\sigma$, once the three coupling constants 
$g_{\omega N}$, $g_{\rho N}$ and $g_{\sigma}^{' B}$ are adjusted, the other
coupling constants of the hyperons to the meson fields can be obtained
by employing the SU(6) symmetry from Eqs. (6) and (7). For this purpose, 
the QMC model is solved for symmetric nuclear matter, 
and as in Ref. \cite{Jin2}, for a given value of $g^q_\sigma = 1$
the coupling constants $g_{\omega N}$, $g_{\rho N}$ and $g_{\sigma}^{' B}$ 
are adjusted to reproduce the nuclear matter binding energy $B/A = 16$ MeV
at saturation density $n_0 = 0.17$  fm$^{-3}$ and symmetry energy 
$a_{\rm sym} = 32.5$ MeV. The resulting coupling constants are given in
Table II. For the parametrization employed here, the predicted values of
effective nucleon mass and compressibility at saturation density are
$m_N^*/m_N = 0.78$ and $K = 289$ MeV. It may be worth mentioning that in
order to reproduce the same saturation properties of density, binding 
and symmetry energies, the coupling constants required in the original QMC
model with bag fixed at $B^{1/4} \equiv B^{1/4}_0 = 188.1$ MeV are
$g_\sigma^2/4\pi = 20.2$, $g_\omega^2/4\pi = 1.55$ and $g_\rho^2/4\pi = 5.51$
with a relatively larger effective mass $m_N^*/m_N = 0.89$ and smaller 
compressibility $K = 220$ MeV. The higher effective mass and thereby smaller
scalar field potential is compensated at the saturation density by a 
smaller vector field i.e.,
a smaller coupling $g_\omega^2/4\pi$. Since at high densities the vector field
dominates over the scalar, the smaller vector coupling leads to a softer EOS. 
The parameters obtained here are entirely from free space value and from nuclear
matter at the saturation density, therefore this set can be used also in the 
model with the two strange mesons. In the following we refer to the model where
the interaction is mediated by ($\sigma$, $\omega$, $\rho$) mesons as QMCI
while its extension by incorporating ($\sigma^*$, $\phi$) mesons as QMCII.

\subsection{The Extended Quantum Hadrodynamics Model}

Quark meson coupling models are designed to describe both the bulk properties
of nuclear systems and medium modifications of the internal structure of the
baryon. Before any reliable predictions for changes due to the quark 
substructure can be made especially at large 
densities relevant to the core of neutron stars,
it is important that the QMC model predicts the established results of nuclear
phenomenology obtained in the quantum hadrodynamics model (QHD) near the 
saturation density. In QHD the relevant degrees of freedom are the structureless
baryons interacting by the exchange of mesons. The direct coupling of the bag
constant to the scalar field in the QMC model mimics scalar self-interaction 
terms. Therefore, for a consistent comparison with this QMC model, we employ a
version of QHD model which contains the cubic and quartic scalar 
self-interactions \cite{Bog}. The Lagrangian for the baryon octet in the 
QHD model within the RMF approximation is similar to that of Eq. (13) 
for the QMC model, but for the baryonic effective mass given by
\begin{equation}
m_B^* = m_B -  g_{\sigma B} \sigma -  g_{\sigma^* B} \sigma^*  ~,
\end{equation}
and contains a scalar self-interaction term
\begin{equation}
U(\sigma) = \frac{g_2}{3} \sigma^3 + \frac{g_3}{4} \sigma^4 ~, 
\end{equation}
proposed by Boguta and Bodmer \cite{Bog} to get a correct compressibility at 
saturation density. Comparing Eqs. (14) and (27), it is clear that the 
coupling constants
in QHD are independent of the scalar field and they are determined at the 
saturation density. In fact it has been demonstrated \cite{Mul97} that QMC 
model is formally equivalent to the nuclear QHD model with a field dependent 
scalar-nucleon coupling. The equations of motion for only the scalar meson 
fields of Eqs. (15) and (16) are then modified in the QHD model to
\begin{equation}
m_\sigma^2\sigma + \frac{\partial}{\partial\sigma}U(\sigma) 
= \sum_B g_{\sigma B} \frac{2J_B + 1}{2\pi^2}
\int_0^{k_B} \frac{m_B^*}{\left[k^2 + m_B^{* 2}\right]^{1/2}} \: k^2 \ dk ~,
\end{equation}
\begin{equation}
m_{\sigma^*}^2\sigma^* = \sum_B g_{\sigma^* B} 
\frac{2J_B + 1}{2\pi^2} \int_0^{k_B} \frac{m_B^*}
{\left[k^2 + m_B^{* 2}\right]^{1/2}} \: k^2 \ dk ~.
\end{equation}
Here nonlinear interaction only for the $\sigma$ meson is employed, the
interaction between the hyperons in the QHD model are through linear 
$\sigma^*$ and $\phi$ mesons. The fact that all the other meson field 
equations are unaltered in QHD, suggests that in the QMC model the effect of
the internal quark structure of a baryon enters entirely through the factor
$C_B(\sigma)$ and $C_B(\sigma^*)$ as has been mentioned in Ref. \cite{Sai94}.
The vector fields in QMC cause only a shift in the quark wave functions. The 
equations of motions for the meson fields in QHD are solved self-consistently 
in accordance with the charge neutrality and $\beta$-equilibrium conditions to
obtain the composition and structure of a neutron star. The five coupling
constants $g_{\sigma N}$, $g_{\omega N}$, $g_{\rho N}$, $g_2$ and $g_3$ in
this model are determined by reproducing the same equilibrium properties of
saturation density, binding energy, symmetry energy, effective mass and 
compression modulus of the QMC model; these are given in Table II.
All the other coupling constants can be obtained from these couplings.
The constant $g_3$ in the scalar field potential (28) is found positive.
This avoids the fatal problem for a quantum field theory that the energy
functional may be unbounded from below which leads to instabilities at high
densities with large scalar fields. Hereafter we refer to the QHD model with
($\sigma$, $\omega$, $\rho$) mesons as QHDI and its extension by introducing 
($\sigma^*$, $\phi$) mesons as QHDII.

\section{Results and Discussions}

In this section we shall present results for baryonic matter in charge
neutral and $\beta$-equilibrium conditions appropriate for a neutron star
in the QMC and QHD models. The effective baryon masses $m^*_B/m_B$ defined 
in Eq. (9) are shown in Fig. 1 as a function of baryon density $n_B$ for the 
models QMCI and QMCII. Unless otherwise mentioned the thin lines refer to 
results for different species in the model QMCI, while the thick ones 
correspond to those in the QMCII 
model. The effective masses of the nucleons rapidly decrease with increasing 
density and then saturate at higher densities. Since the nucleons do not couple 
to the strange scalar field $\sigma^*$, their masses in the models I and II 
are identical. At densities around nuclear matter values where hyperons have 
not appeared, the effective mass of a test hyperon is determined by only the 
scalar field $\sigma$ created by nucleons which is assumed to be 
unaffected by inserting the hyperon. Consequently, the effective masses of 
all the baryons for $n_B \leq 2n_0$ reveals no splitting in models I and II.
Moreover, the different baryonic masses at the saturation 
density indicates the SU(6) symmetry (based on the number of light quarks 
counting rule) for $\sigma-B$ coupling i.e., 
$\delta m^*_{\Lambda,\Sigma}/\delta m_N^* = 2/3$ and 
$\delta m^*_\Xi/\delta m_N^* = 1/3$. The QMCI model respects this scaling
relation to nearly the entire density range explored here. On the other
hand, in the QMCII model at densities $n_B \approx 0.38$ fm$^{-3}$ when the 
hyperons ($\Sigma^-$, $\Lambda$) production threshold is reached (see Fig. 6), 
the attractive scalar field $\sigma^*$ starts to contribute. The reduction 
in the mass of strange quark in model II entails a substantial decrease in 
the masses for the hyperons in accordance with the SU(6) relation for 
$\sigma^*-B$ coupling. Since the $\Xi$ hyperon has two strange quarks, its
mass is maximally suppressed by the $\sigma^*$ field. Note that although 
$m^*_N/m_N$ reaches small values at high densities in presence of hyperons, 
it never becomes negative in the density range studied here. In fact, 
relativistic mean field models fitted to the bulk properties of nuclear 
matter with a high $m^*_N/m_N\approx 0.7-0.82$ lead to a finite effective 
mass even at central densities of maximum mass stars. However, the 
effective masses in this case become negative only at densities much higher
than the central densities of maximum mass stars
\cite{Kno}. In principle parameter sets fitted to
finite nuclei properties can also be obtained. However, in most of these 
sets the effective masses of nucleons in presence of hyperons get negative 
at densities much smaller than that of maximum mass stars.
Therefore, these sets are not reliable to calculate
neutron star properties. Recall that the original QMC model with bag 
constant fixed to the free space value, leads to much smaller scalar field
and higher effective nucleon mass $m^*_N/m_N = 0.89$. To reproduce the correct
spin-orbit splitting, the reduction in the effective mass ($m^*_N/m_N = 0.78$)
is achieved by direct coupling of the bag constant to the scalar fields.

The variation of bag constant for the baryons $B_B/B_0$ (see Eq. (11)) with
density is shown in Fig. 2 for the QMCI and QMCII models. The bag constants
decrease with density and saturate at high densities. This behavior is 
similar to that of effective baryon masses as they have been obtained in a 
self-consistent manner. The decrease of bag constant relative to its free 
space value implies a decrease of bag pressure which causes an 
increase of bag radius in the medium.
The variation of the bag radius $R_B/R_0$ (relative to its free space value
$R_0$ which is different for different baryons; see Table I) with density
is shown in Fig. 3. At saturation density when the bag constant for nucleon 
has decreased to $B_N/B_0 = 0.45$, the corresponding radius has increased to 
$R_N/R_0 = 1.22$. At densities of $n_B = (6-8)n_0$ corresponding to the
maximum mass of neutron stars, the nucleon bag constant has decreased to a 
significantly small value $B_N/B_0 = (0.093 - 0.065)$  while the corresponding
radius is $75 - 88 \%$ larger than its free space value. This implies a
considerably swollen nucleon (and hyperons) in the star matter; a detailed
discussion of its consequences is given later.

We now present a systematic comparison between the QMC and the QHD models
for neutron star matter with their coupling constants determined to 
reproduce the same set of nuclear matter saturation properties as given in 
Table II. In Fig. 4, the baryon effective masses $m^*_B/m_B$ as a function
of density $n_B$ are displayed for the models QHDII (thin lines) and QMCII
(thick lines). It is seen that at low and moderate densities, the $m_B^*$'s
for the two models are in good agreement, this is not surprising as both 
the models are calibrated 
to the same properties at nuclear matter densities. At higher densities,
especially when hyperons start to populate (at $n_B \approx 0.38$ fm$^{-3}$), 
the effective masses are rather distinct in the two models. The pure scalar
and vector field strengths are shown in Fig. 5 as a function of density for
the models QMCII (top panel) and QHDII (bottom panel). The potentials for a
given baryon species are obtained by multiplying them with the corresponding 
coupling constants listed in Table II; for the scalar fields these couplings
are however density dependent for QMCII. A careful examination of Fig. 5 
indicates that the values of the fields and potentials for $\omega$ 
in the two models are nearly identical over the entire density range. 
On the other hand, at all densities the $\sigma$ field in QMCII is larger 
than that in QHDII. However, the decreasing coupling constant 
$g_{\sigma N}(\sigma)$ with $n_B$ in the former model causes 
the potential $U_\sigma = g_{\sigma N}(\sigma)\sigma$ to be the same as 
QHDII at the normal nuclear matter density. This lead to the same saturation
properties (binding energy and density) in the two models. At densities 
higher than the normal nuclear matter value, $g_{\sigma N}(\sigma)$ 
further decreases causing the potential $U_\sigma$
in QMCII to saturate earlier than QHDII. In other words, the scalar 
density factor $C_N(\sigma)$ [see Eqs. (15) and (20)] in QMCII decreases 
with increasing $g^q_\sigma \sigma$ (or $n_B$) as quarks in the medium 
become more relativistic \cite{Sai94,Tsu}. As a consequence, the drop in the 
nucleon effective mass relative to its free space value in QMCII is smaller 
than that in QHDII. Clearly at high densities the quark substructure of the 
nucleon plays a crucial role in QMC model. This feature of larger effective 
masses in QMCII is more evident for the hyperons. This is because of a smaller 
value of the attractive scalar field $\sigma^*$ (see Fig. 5) and a
decreasing in-medium coupling constant $g_{\sigma^* B}(\sigma^*)$ 
results in a much smaller potential 
$U_{\sigma^*} = g_{\sigma^* B}(\sigma^*)\sigma^*$ for QMCII compared to
that in QHDII. The effect is most pronounced for $\Xi$ having two 
strange quarks. We have observed similar qualitative
differences in models I, i.e. between QMCI and QHDI. However, the distinction
in $m_B^*$ for hyperons in the two models is not so profound as they lack
the $\sigma^*$ meson. This indicates the importance of the strange mesons
($\sigma^*, \phi$) which helps in revealing more clearly the quark structure 
of the baryons at high densities in the QMCII model as compared to the 
structureless baryons inherent in the QHDII model.

Having investigated the crucial role played by quarks confined within the 
baryons at high densities, we shall now consider whether the 
QMC model based on nonoverlapping bags can at all be extended to densities
appropriate to neutron star interior. As is evident from Fig. 3, at rather
large densities the increasing bag radius implies considerable overlapping
between the bags and the nonoverlapping bag picture of nuclear matter may break
down because the effects of short-range correlations among the quarks which
should be associated with the overlap of hadrons are {\it explicitly} neglected
in the quark-meson coupling model. We however observe that in the present 
QMC model the physical observables for instance the effective masses, fields 
and therefore the equations of state (see Fig. 7) indicate a smooth and 
continuous behavior from low to very high densities without any dramatic 
discontinuity. Any deviation at the high density regime from the QHD results 
may be interpreted as the interesting effects due to quark structure of the
baryons only. This is in consonance with the argument put forward \cite{Bro} 
that the physical observables do not depend on the bag radius. On
contrary, in the original QMC model with bag fixed at the free space value,
the observables were found \cite{Mul97} to differ drastically from QHD results
with large discontinuities. Moreover, solutions in that version of QMC model 
cease to exist above $n_B \approx 4.92 n_0$ since the eigenvalue $x_q$
in Eq. (2) vanishes. This indicates that by using a density dependent bag 
constant through direct coupling to the scalar field, the QMC model not only 
reproduces the correct spin-orbit potential but possibly also includes the 
effects of quark-quark correlations associated with overlapping bags which was 
missing in the original QMC model. A possible explanation to this may follow 
from an alternative approach inspired by effective field theories. In this 
theory the constraint of renormalizability of the hadronic Lagrangian is 
abandoned \cite{Fur96,Mul96} which allows to introduce higher order meson 
self-interaction (i.e. orders higher than the quartic scalar interaction) 
consistent with the underlying symmetries of QCD. 
By a suitable truncation at some low orders of the fields,
it was argued \cite{Fur96} that by allowing non-linearities in the meson fields 
that generate additional density dependence in the interactions, the important
effects of correlations (and exchange) between the nucleons beyond the simple
Hartree contributions are automatically included. Thus, one can include many
body effects beyond the simple Hartree level even though only classical meson
fields and local interactions are retained. 

In the context of QMC model, as demonstrated in this paper and also in 
Ref. \cite{Mul97}, an exponential dependence of bag constant $B_B$ on $\sigma$ 
(and $\sigma^*$) is equivalent to scalar self-interactions of infinite 
orders, of course each term is smaller by a factor 
$\sim (g^{' B}_\omega/m_B)$ than its preceding one [see Eq. (11)]. Hence these 
higher order self-interaction terms possibly include (implicitly) the effects 
of quark-quark correlations at high densities providing a non-discontinuity 
in the physical observables, unlike the original version of the QMC model with
constant bag. This suggests that in a QMC model with bag constant coupled
to the scalar field, one can in principle extrapolate to high densities to
explore the neutron star properties.

\section{Composition and Structure of Neutron Star Matter}

In this section the constitution and structure of stable charge neutral
(neutron star) matter in the supernuclear density regime are presented in the
QMC and QHD models. In Fig. 6 the abundances of baryons and leptons as a
function of density in the star matter are shown for the models QMCII 
(top panel) and QHDII (bottom panel). 
At densities slightly below the nuclear matter value,
the $\beta$-decay of neutrons to muons are allowed, and thus muons start to 
populate. The charge neutrality of a star forces a high isospin asymmetry
so both the electron chemical potential $\mu_e$ and the $\rho$ field,
$\rho_{03}$, grow at low density as evident from Fig. 5. Although the $\rho$
field is very small and never exceeds $-18$ MeV, its correct determination
from symmetry energy is of utmost importance as it determines the proton
fraction. It was demonstrated \cite{Lat} that for a $npe$ system 
rapid cooling by nucleon direct URCA process is allowed by the momentum 
conservation condition $k_p + k_e \geq k_n$ which corresponds to a proton 
fraction $Y_p \geq 0.11$. In both the models used here this condition is 
satisfied at densities $n_B \geq 0.28$ fm$^{-3}$ thus rapid cooling by direct
URCA process can occur. In the absence of any hyperons, the charge neutrality 
condition forces the proton (and lepton) fractions to continuously increase 
with density in these relativistic mean field models. Thus once the threshold
density for cooling by direct URCA process is achieved, it would persist up
to the center of such stars. In contrast, the decreasing symmetry energy at
high densities in the nonrelativistic models would limit cooling near the
central region of massive stars. The symmetry energy also determines the 
hyperon production threshold density obtained by the condition 
$\mu_B = \mu_n - q_B\mu_e \geq \varepsilon_B(k=0)$, where the energy
eigenvalue $\varepsilon_B$ is given by Eq. (24). Consequently the threshold
density for a hyperon species is determined by its charge and effective mass 
and by all the fields present in the system.
As expected, the $\Lambda$ with mass 1116 MeV and $\Sigma^-$ 
with a mass 1193 MeV appear at roughly the same density $n_B \approx 0.38$
fm$^{-3}$, because the somewhat larger mass of 
$\Sigma^-$ is compensated by its negative charge. Since charge neutrality 
can now occur more economically by $\Sigma^-$, the lepton fraction begins to
fall. The electron chemical potential (see Fig. 5) then saturates around
200 MeV and subsequently decreases with increasing $\Sigma^-$ population.
More massive and positively charged particles than these appear at high 
densities. The substantial reduction in the effective mass of $\Xi$ in QHDII 
(as shown in Fig. 4) is manifested by a relatively early appearance of $\Xi^-$
and $\Xi^0$ and their larger abundances in the star compared to that in QMCII. 
The enhanced $\Xi^-$ production in turn causes a further rapid 
decrease of the lepton fraction in QHDII. At high densities, all the baryons
tend to saturate with the abundance of $\Lambda$ being maximum, even exceeding
the number of neutrons. Because of the fast growth of the hyperons and their 
comparable abundances with the nucleons, the dense interior of a star resembles
more a hyperon star than a neutron star. The net strangeness fraction for stars
in the model QHDII is slightly enhanced due to larger $\Xi$ abundance 
than that in QMCII.
In contrast to pure $npe$ stars, the proton fraction
here reaches maximum value once the hyperons ($\Sigma$, $\Lambda$) start 
populating and thereafter it saturates at the level of $20\%$. Therefore rapid
cooling by nucleon direct URCA process can still occur in these stars. Since
the critical density of nucleon direct URCA process is nearly identical to
the hyperon threshold density, and the emmisivities from the hyperon direct
URCA processes are about $5-100$ times smaller than that from the nucleons
\cite{Pra92}, the stars cooling by direct URCA process is dominated by 
nucleons $-$ the hyperons would only have a minor contribution to it.
In the models QMCI and QHDI, the effective masses of the hyperons and 
the potentials are however found to be almost identical, 
hence the composition of the stars in model I were found to be practically 
indistinguishable.

The equation of state, pressure $P$ versus the energy density $\varepsilon$
is displayed in Fig. 7 for the different models studied here. The EOS for
nucleons only (np) star shown for the QMC model (solid line) is found 
to be considerably stiff. Since the corresponding EOS in the QHD model is 
found to be nearly identical we do not present the result for clarity. At high 
densities when the Fermi energy of nucleons exceeds the effective mass of 
hyperons minus their associated interaction energy, the conversion of nucleons 
to hyperons is energetically favorable. Since this conversion relieves the 
Fermi pressure of the nucleons, the equation of state is softened. 
This effect is further accentuated by the decrease of the pressure exerted by 
leptons because of their replacement by negatively charged hyperons in 
maintaining charge neutrality more efficiently. The EOS for nucleons plus 
hyperons (npH) system for both the models I and II are shown in the figure; the
thick lines refer to QMC results while the thin lines correspond to QHD.
Several structures observed in the equation of state correspond to the 
densities at which different hyperon species begins to populate. In model I
(shown by dashed lines), the EOS in QMCI is found to be softer compared to
QHDI. In this situation the vector fields ($\omega$, $\rho$) and the 
effective hyperon masses are found to be same in the two models, only the 
scalar field $\sigma$ and the effective mass of nucleons
in QMCI is somewhat larger than QHDI. This results in the QMCI a larger 
contribution from the scalar attraction and a smaller one from the repulsive 
kinetic term of nucleons to the pressure (Eq. (26)) leading to a softened EOS. 
With the inclusion of strange mesons, the effective masses of hyperons also
undergo significant reduction (see Fig. 1) while, in general, the contribution
from the repulsive $\phi$ field dominates over the attraction from $\sigma^*$
field. The net effect is thus a stiffer EOS in model II (shown by dash-dotted
lines) in contrast to model I. As evident from Figs. 4 and 5, the combined
effects of enhanced effective masses, considerably large repulsive $\phi$ field
and smaller attraction from $\sigma^*$ field act in increasing the pressure
in QMCII compared to QHDII. This entails a pronounced stiffening with the EOS 
for QMCII being even stiffer relative to QHDII. Thus by including the strange 
mesons, a complete reversal in behavior occurs for the equation of state in 
QMC and QHD models which should have a significant
bearing on the structure of the stars. For comparison, the causal limit 
$p=\varepsilon$ is also shown in the figure. All the relativistic models
studies here respect causality condition $\partial p/\partial\varepsilon \leq 1$
so that the speed of sound remains lower than the speed of light.

An important parameter describing the equations of state is the adiabatic index
$\Gamma = d \ln P/d \ln n_B = 
(P + \varepsilon)/P \cdot dP/d\varepsilon$.
In Newtonian theory it is possible to find a stable hydrostatic configuration
for a spherical mass distribution if the adiabatic index exceeds $4/3$ 
\cite{Cha69}, and when general relativity is included it slightly exceeds
this value \cite{Coo}. Fig. 8 shows the adiabatic index $\Gamma$ versus
the energy density $\varepsilon$ for different equations of state. At 
densities below the nuclear matter value, $\Gamma$ could have very small values
since most the pressure support for the star originates from the electrons, 
the nuclear phase has a negative contribution to pressure. At densities 
greater than about $n_0$, the EOS stiffens and $\Gamma$ is significantly 
greater than $4/3$. With the appearance of hyperons, the softening of the 
EOS is manifested by the considerable lowering of the adiabatic index. The 
several structures observed in the EOS corresponding to the population of 
hyperons at the threshold densities is clearly evident in Fig. 8. 
The adiabatic index drops at each density when a new hyperon species is 
populated.

The differences in the EOS at high densities are expected to be reflected
in the structure of the neutron stars, namely their masses and radii. The 
static neutron star sequences representing the stellar masses $M/M_\odot$
and the corresponding central energy densities $\varepsilon_c$ obtained
by solving the Tolman-Oppenheimer-Volkoff equations \cite{Tol} are shown
in Fig. 9 for different equations of state. In general, such a sequence
posses a minimum mass below which gravitational attraction is not 
sufficient against the radial oscillations that destroy these configurations 
by dispersal. On the other hand, a maximum mass of the sequence exits beyond 
which the pressure support from the EOS is insufficient against the strong 
gravitational attraction. Stars beyond this mass are unstable to acoustical 
radial vibrations and thereby collapse to a black hole. The crustal region 
has a negligible contribution ($\sim 10^{-5}M_\odot$) to the total 
mass of a star, while most of the mass originates from the dense interior
beyond the saturation density. Thus mass measurements may provide considerable
insight into the interior constitution of a star. For the np system, the
extremely stiff EOS corresponds to a large Fermi pressure and hence can sustain
large limiting mass. The maximum masses for such np stars in the QMC model
is $M_{\rm max} = 1.988M_\odot$, while a relatively softer EOS in the 
corresponding QHD model results in $M_{\rm max} = 1.962M_\odot$. In Table III,
the maximum masses and the corresponding radii $R_{M_{\rm max}}$ and central 
baryonic densities $n_c$ are presented. 
In Fig. 10, we also show the mass-radius relationship of the
different EOS. With the inclusion of more baryon species in the form of 
hyperons, the considerable softening of the EOS results in relatively
much smaller mass stars. Since the QMCI model has a much softer EOS than
QHDI, the $M_{\rm max}$ values are $1.478M_\odot$ and 
$1.488M_\odot$ respectively; the star sequence and the mass-radius 
relationship in these 
models are not shown in the figures for clarity. The larger radii obtained
in stars with hyperonization are a consequence of weaker gravitational 
attraction from the smaller masses that causes the stars to be large and
diffuse. Compared to the mass of a star where the contribution is primarily from
beyond the saturation density, about $40\%$ of the star's radius originates
from the EOS at $n_B \stackrel{<}{\sim} n_0$. Consequently, radius
measurements should be rather insensitive to the changes in the EOS due to
quark structure of baryons or to the interior constitution of the star. 
In fact, no precise radius measurements currently exist. The star sequences 
and the mass-radius relationships in model II are shown in Figs. 9 and 10,
respectively. The reversal in behavior observed in the EOS with the addition 
of strange mesons, i.e. the EOS is stiffer in QMCII than QHDII, is 
manifested in the maximum masses and the corresponding radii for such stars. 
For all the cases studied here, the maximum masses of the stars are found
to be larger than the current observational lower limit of $1.44M_\odot$ 
imposed by the larger mass of the binary pulsar PSR 1913 + 16 \cite{Wei}.
When hyperons are included, the central densities reached for all the stars 
above this lower mass limit suggest that the inner cores of these stars are 
rich in hyperons.

Constraints on the high density EOS can be derived from the measurement of 
the absolute upper limit to the rotation velocity $\Omega_{\rm max}$ of a
neutron star. This is possible because the maximum angular velocity of 
rotating neutron stars, in general, is an increasing function of the
softness of the EOS which corresponds to more compact stars with high 
central densities and smaller equatorial radius. The maximum rotation
rate of a neutron star is determined by the condition that the equatorial
surface velocity equals the Keplerian velocity $-$ the orbital velocity of a
particle at the equator. At the Keplerian frequency, the rotating star is
unstable with respect to mass shedding from the equator. In reality, the
rotation may be more severely limited by gravitational radiation instability
to nonaxisymmetric perturbations \cite{Cha70}. However, this instability has
been later shown to be stabilized by the existence of the viscosity of stars
at homogeneous density \cite{Lin}. Therefore, the Keplerian rate may be 
considered as a reasonable estimate of the maximum rotation rate of a neutron
star. The calculation of the Keplerian velocity for a given EOS is quite
involved and has to be performed with full general relativity. On the other
hand, a precise universal empirical formula was found by Haensel and Zdunik
\cite{Hae} for the maximum angular frequency $\Omega_{\rm max}$ for rigid 
rotation in terms of maximum mass and radius of a nonrotating star for a 
given EOS: $\Omega_{\rm max} = {\cal C}_\Omega (M_{\rm max}/M_\odot)^{1/2} 
(R_{M_{\rm max}} /10 \ {\rm km})^{-3/2} \ {\rm s}^{-1}$, where $M_{\rm max}$ 
and $R_{M_{\rm max}}$ are the maximum mass and corresponding radius of the 
nonrotating star. The 
dimensionless phenomenological constant independent of the EOS was found 
\cite{Hae} to be ${\cal C}_\Omega = 7750$. Later calculations within the full 
framework of general relativity have shown \cite{Fri92} that this formula 
has an accuracy better than $5 \%$ for a wide range of realistic EOS 
which are both causal
and stiff enough to support observed maximum mass $M_{\rm max} = 1.44 M_\odot$.
Since rotational energy stabilize a star, the most massive rotating star has 
more mass and radius than the maximum mass and radius of the corresponding 
nonrotating star. For our present equations of state, the Keplerian frequencies
obtained by using the above formula are given in Table III for the respective
rotating stars. It is observed that $\Omega_{\rm max}$ closely follow the
trend of the masses (and radii) of different models with the nucleons only 
star having the largest rotation rate and stars with hyperons in model I are
the slowest. As mentioned above, this may be attributed to the smallest mass
and thereby largest radius due to small gravitational attraction in the 
softest equation of state and vice versa.

\section{Summary and Conclusion}

In this paper we have investigated the effects of the internal quark structure
of baryons on the neutron star properties within the relativistic mean-field
quark-meson coupling model. This model describes baryons as nonoverlapping MIT
bags in which the quarks interact through scalar and vector mean fields.
Before any reliable prediction of the quark structure effects on neutron star
properties at high densities can be made, it is essential that the QMC model
reproduces at around nuclear matter densities the results of more established
quantum hadrodynamics model where the relevant degrees of freedom are the 
structureless baryons. In the original QMC model, the bag constant was fixed
at the free space value as a consequence of which the model predicts much 
smaller attractive scalar potential and hence smaller spin-orbit potential
compared to the experimental results and that obtained in the QHD model. By
considering a medium (density) dependent bag constant parametrized through
direct coupling to the scalar field, correct spin-orbit splitting was observed.
Moreover, this medium modified QMC model is found to be in excellent agreement
with the low and moderate density results of QHD model with a general nonlinear
scalar potential when both the models are calibrated to produce the same nuclear
matter saturation properties. This improved fit is obtained by employing a bag 
constant which is a decreasing function of density, which however implies an 
increasing bag radius. We are therefore faced with the problem that due to 
the increasing bag radius and thereby overlapping bags the QMC model may 
not be applicable at the high density regime relevant to central densities 
of massive stars. A natural test of the reliability of the model at 
high densities then lies in the fact that the equation of state should be 
well-behaved and continuous when extrapolated to the extremes of density. 
Indeed, we have found that the physical observables exhibit reasonable 
behavior without any discontinuity at high densities up to $n_B \approx 10n_0$ 
studied here. The deviation of the results
observed at high densities in the two effective field theoretical models,
the QMC and QHD, with different underlying basic constituents may be interpreted
as primarily arising from the crucial effects of the quark structure. The 
original version of the QMC model, where the important effect of quark-quark
correlation associated with overlapping bags is neglected, exhibited 
discontinuity and therefore could not be extrapolated to high densities. A 
direct coupling of the bag constant to the scalar field is equivalent to higher
powers of (nonlinear) scalar interactions. According to the modern viewpoint
of effective field theory, higher order meson interaction includes the effect
of correlations. Therefore, by employing a medium (scalar field) dependent
bag constant, we not only could reproduce the correct scalar potential but 
could also mimic the quark-quark correlations leading to a well-behaved EOS at 
high densities.

We have included two additional (hidden) strange mesons which couple only to 
strange quarks in a baryon bag in the QMC model and to hyperons in the QHD 
model. The rather strong hyperon-hyperon interaction can be accounted by these
mesons. The coupling constants of the quarks and hyperons have been fixed by 
SU(6) symmetry relations based on quark-counting argument. The strange mesons
are found to have considerable influence on the composition and structure of
neutron star matter with hyperons. In absence of these mesons, the model QMCI
exhibits softer EOS with smaller maximum mass star and larger corresponding 
radii than that in QHDI. With the inclusion of strange mesons, the additional 
attraction
imparted by the scalar meson $\sigma^*$ causes a drop in the effective masses
of hyperons, the decrease being determined by the number of strange quarks in 
the baryons. However, the two mesons ($\sigma^*$, $\phi$) has an overall 
repulsive effect so that the EOS in the model II is stiffer compared to 
model I without the strange mesons. The repulsion is maximum in QMCII because 
of the decreasing in-medium scalar-baryon coupling constant and smaller 
scalar field $\sigma^*$ than in QHDII. Consequently, the EOS in QMCII is 
significantly stiffened (and even stiffer than QHDI) with relatively larger 
maximum mass star and corresponding smaller radius. As observed in previous 
studies, the EOS is found to be considerably softened by incorporating hyperons
as the new degrees of freedom which appear at $n_B \approx 2n_0$. The center 
of massive stars are found with comparable abundance of hyperons and nucleons; 
the strangeness fraction of stars in QMC models are relatively higher than in 
QHD models because of enhanced $\Xi$ production. Rapid cooling by direct URCA 
process of all these stars are found to be dominated by nucleons due to large 
proton fraction ($Y_p \geq 0.15$), the hyperons add only $\sim 5\%$ to this 
more dominant process. It therefore seems difficult to differentiate stars 
in the different models studied here with and without hyperons by rapid 
cooling procedure.
\vspace{1cm}

\noindent Acknowledgment: S.P. gratefully acknowledges support
from the Alexander von Humboldt Foundation and Institut f\"ur Theoretische 
Physik, J.W. Goethe-Universit\"at for kind hospitality.


\newpage

\begin{table}

\caption{The free space values of bag parameters $z_0$ and bag radii $R_0$ 
for different baryons obtained by reproducing the baryonic masses $m_B$ 
in free space. The bag constant for the baryons at the free space value 
is $B_0^{1/4} = 188.1$ MeV while the masses of the quarks are taken as 
$m_u = m_d = 0$ and $m_s = 150$ MeV. The strangeness $S_B$ of the baryons
are also given.}

\begin{tabular}{ccccc} 

Baryon& $m_B$ (MeV)& $S_B$& $z_0$& $R_0$ (fm) \\ \hline
$N$& 939& 0& 2.030& 0.600 \\
$\Lambda$& 1116& -1& 1.815& 0.642 \\
$\Sigma$& 1193& -1& 1.629& 0.669 \\
$\Xi$& 1313& -2& 1.505& 0.686 

\end{tabular}
\end{table}
\vspace{.5cm}

\begin{table}

\caption{The coupling constants obtained in the QMC model by reproducing
the saturation density $n_0 = 0.17$ fm$^{-3}$, the binding energy $B/A = 16$ 
MeV and the symmetry energy $a_{\rm sym} = 33.2$ MeV. The coupling constant
of the scalar $\sigma$ field to the bag is $g_\sigma^{' B} = 2.269$. The 
scalar coupling constant corresponds to the free space value [see Eq. (20)] 
while its coupling to the ($u$, $d$) quarks is taken as $g_\sigma^q = 1$. 
The predicted values of compressibility and effective nucleon mass at 
saturation are $K = 289$ MeV and $m^*_N/m_N = 0.78$. The coupling constants 
in the QHD model are obtained by adjusting these same five saturation 
properties. The meson masses are taken to be $m_\sigma = 550$ MeV, 
$m_\omega = 783$ MeV and $m_\rho = 770$ MeV.}

\begin{tabular}{cccccc} 

Model& $g_\sigma^2/4\pi$& $g_\omega^2/4\pi$& $g_\rho^2/4\pi$
& $g_2$ (fm$^{-1}$)& $g_3$ \\ \hline
QMC& 5.184& 5.240& 5.203& $-$& $-$ \\
QHD& 5.174& 5.339& 5.146& 12.139& 48.414

\end{tabular}
\end{table}
\vspace{.5cm}

\begin{table}

\caption{The maximum masses $M_{\rm max}/M_{\odot}$ of nonrotating 
stars and their corresponding radii $R_{M_{\rm max}}$ and central densities 
$n_c$ in the models QMC and QHD. The Keplerian frequency $\Omega_{\rm max}$
for the respective rotating configurations are obtained from the relation
$\Omega_{\rm max} = 7750 (M_{\rm max}/M_\odot)^{1/2} 
(R_{M_{\rm max}} /10 \ {\rm km})^{-3/2} \ {\rm s}^{-1}$. Results are for
stars with only nucleons (np); stars with further inclusion of hyperons (npH) 
in the model I where the interaction is mediated by 
($\sigma$, $\omega$, $\rho$) mesons, and stars in model II where the additional
mesons ($\sigma^*$, $\phi$) are included.}

\begin{tabular}{c|cccc|cccc} 

\hfil& \hfil& QMC& \hfil& \hfil& \hfil& QHD& \hfil& \hfil \\ \hline
\hfil& $M_{\rm max}/M_\odot$& $R_{M_{\rm max}}$& $n_c$& $\Omega_{\rm max}$& 
$M_{\rm max}/M_\odot$& $R_{M_{\rm max}}$& $n_c$& $\Omega_{\rm max}$ \\
\hfil& \hfil& (km)& (fm$^{-3}$)& ($10^3$ s$^{-1}$)&
\hfil& (km)& (fm$^{-3}$)& ($10^3$ s$^{-1}$) \\ \hline
np& 1.988& 10.632& 1.102& 9.969& 1.962& 10.561& 1.110& 10.001 \\
npH (model I)& 1.478& 11.242& 0.965& 7.904& 1.488& 11.092& 0.999& 8.091 \\
npH (model II)& 1.539& 10.823& 1.096& 8.538& 1.491& 11.040& 1.022& 8.159 

\end{tabular}
\end{table}
 \newpage 
\vspace{-5cm}

{\centerline{
\epsfxsize=8cm
\epsfysize=9cm
\epsffile{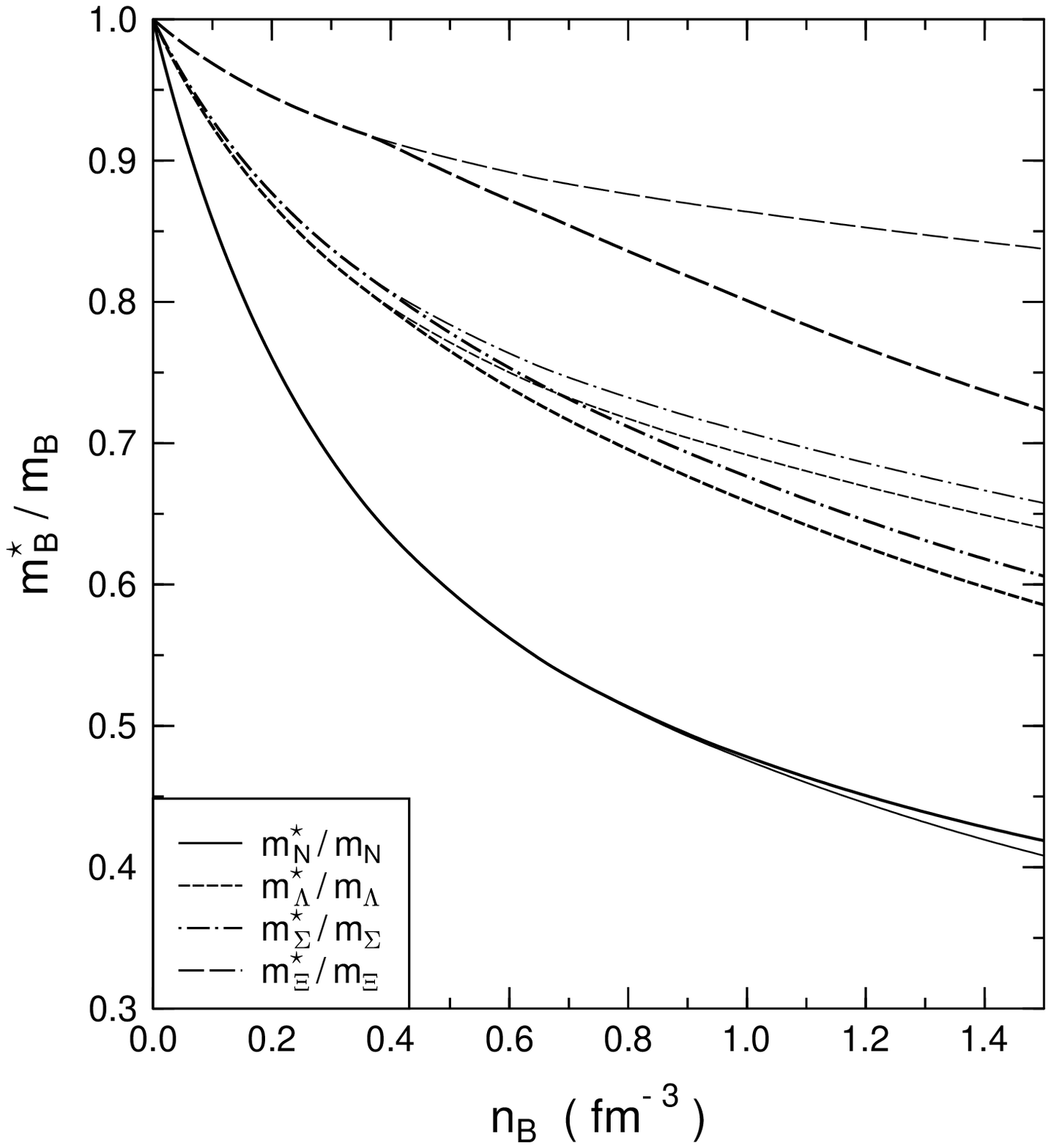}
}}

\vspace{-3.5cm}

\noindent{\small{FIG. 1. The variation of effective  masses $m_B^*/m_B$ of 
the baryons as a 
function of baryon density $n_B$ in the models QMCI (thin lines) 
and QMCII (thick lines)}}.
\vspace{0.7cm}

{\centerline{
\epsfxsize=8cm
\epsfysize=9cm
\epsffile{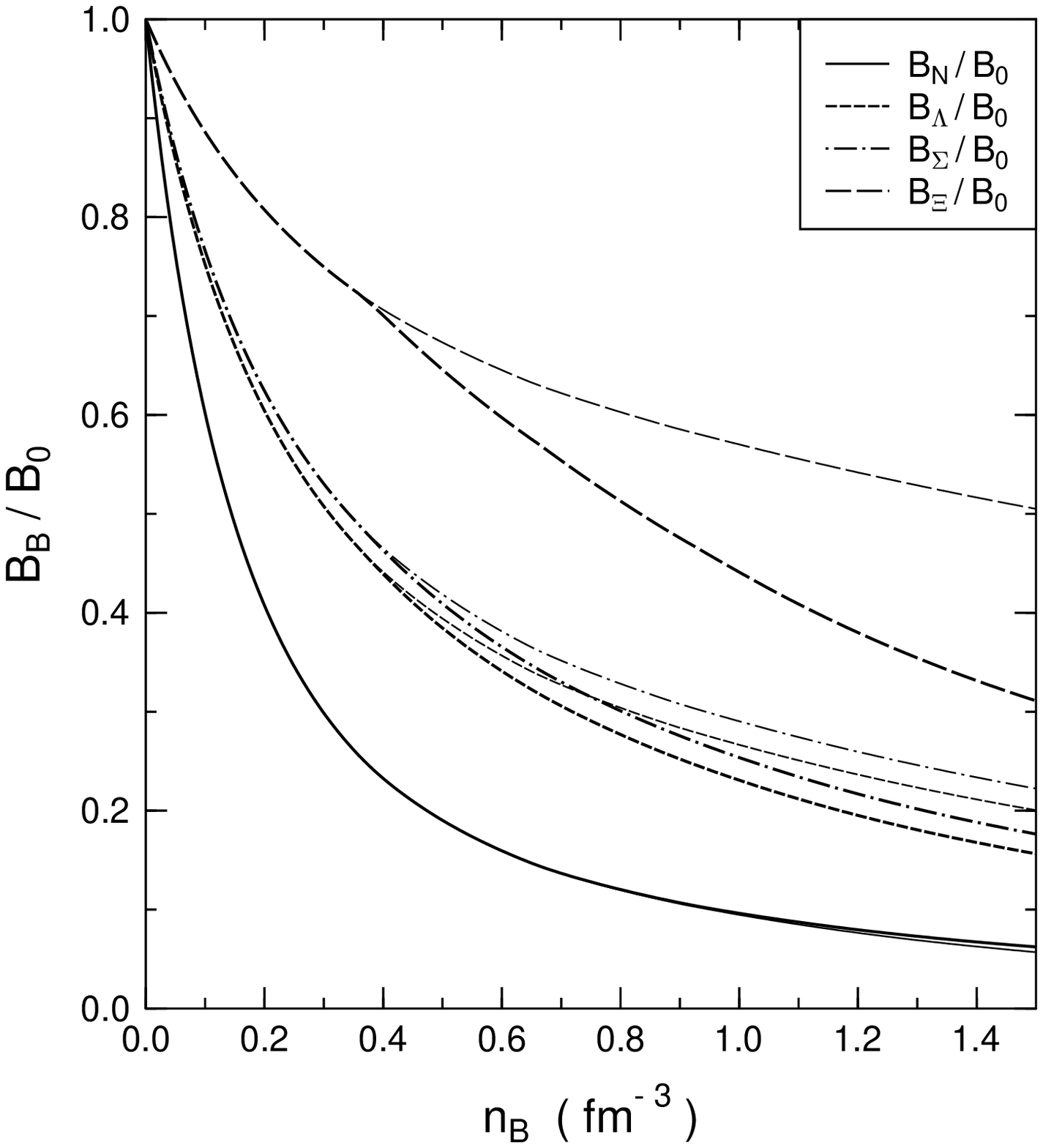}
}}

\vspace{-3.5cm}
\noindent{\small{
FIG. 2. The variation of bag constant $B_B/B_0$ of the baryons as a 
function of baryon density $n_B$ in the models QMCI (thin lines) 
and QMCII (thick lines). The free space bag constant is taken as 
$B_0^{1/4} = 188.1$ MeV.}}
\newpage

\vspace{-5cm}

{\centerline{
\epsfxsize=8cm
\epsfysize=9cm
\epsffile{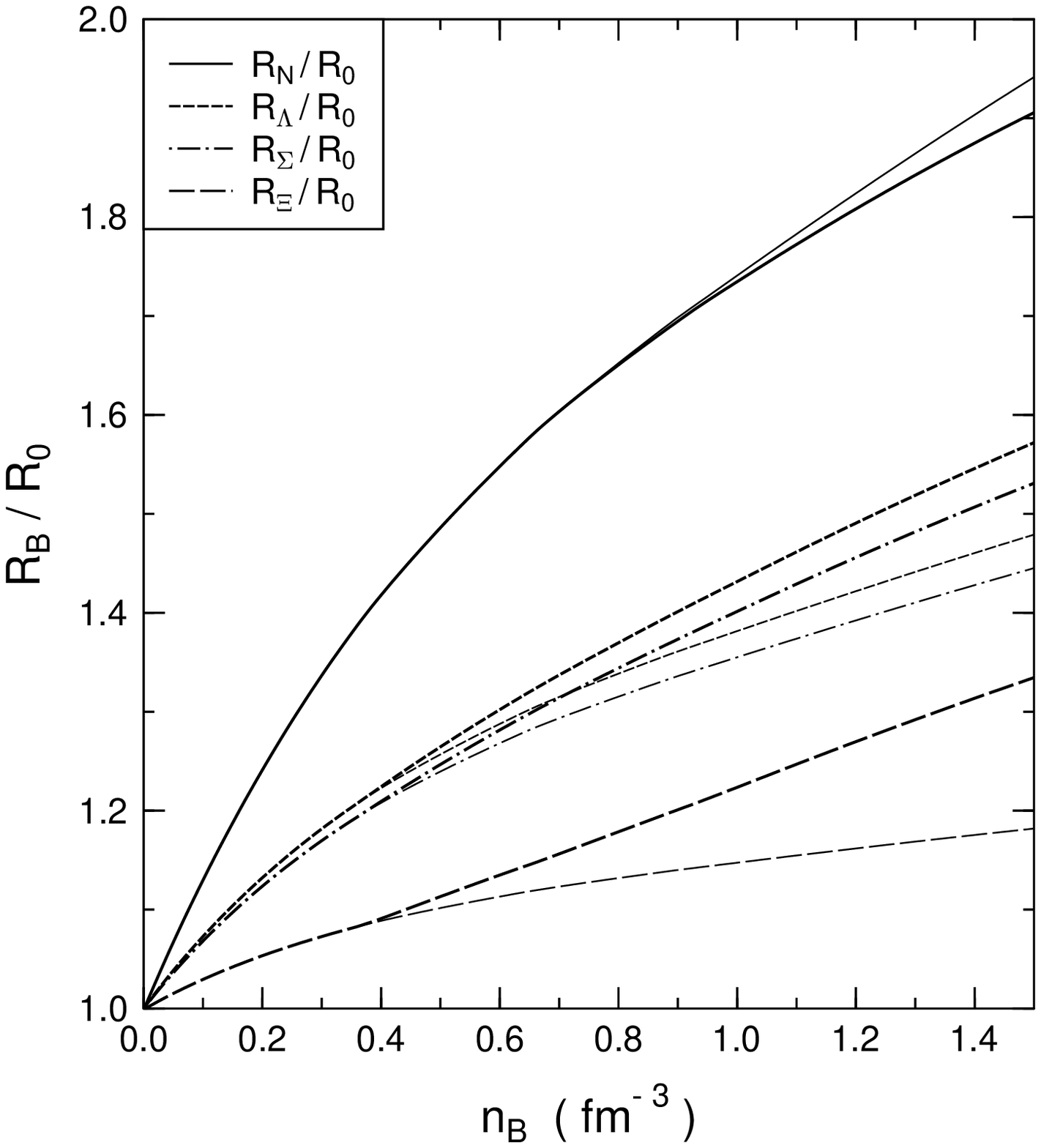}
}}

\vspace{-3.5cm}

\noindent{\small{
FIG. 3. The variation of bag radius $R_B/R_0$ of the baryons as a 
function of baryon density $n_B$ in the models QMCI (thin lines) 
and QMCII (thick lines). The free space bag radius $R_0$ for different 
baryons are given in Table I.}}
\vspace{0.6cm}

{\centerline{
\epsfxsize=8cm
\epsfysize=9cm
\epsffile{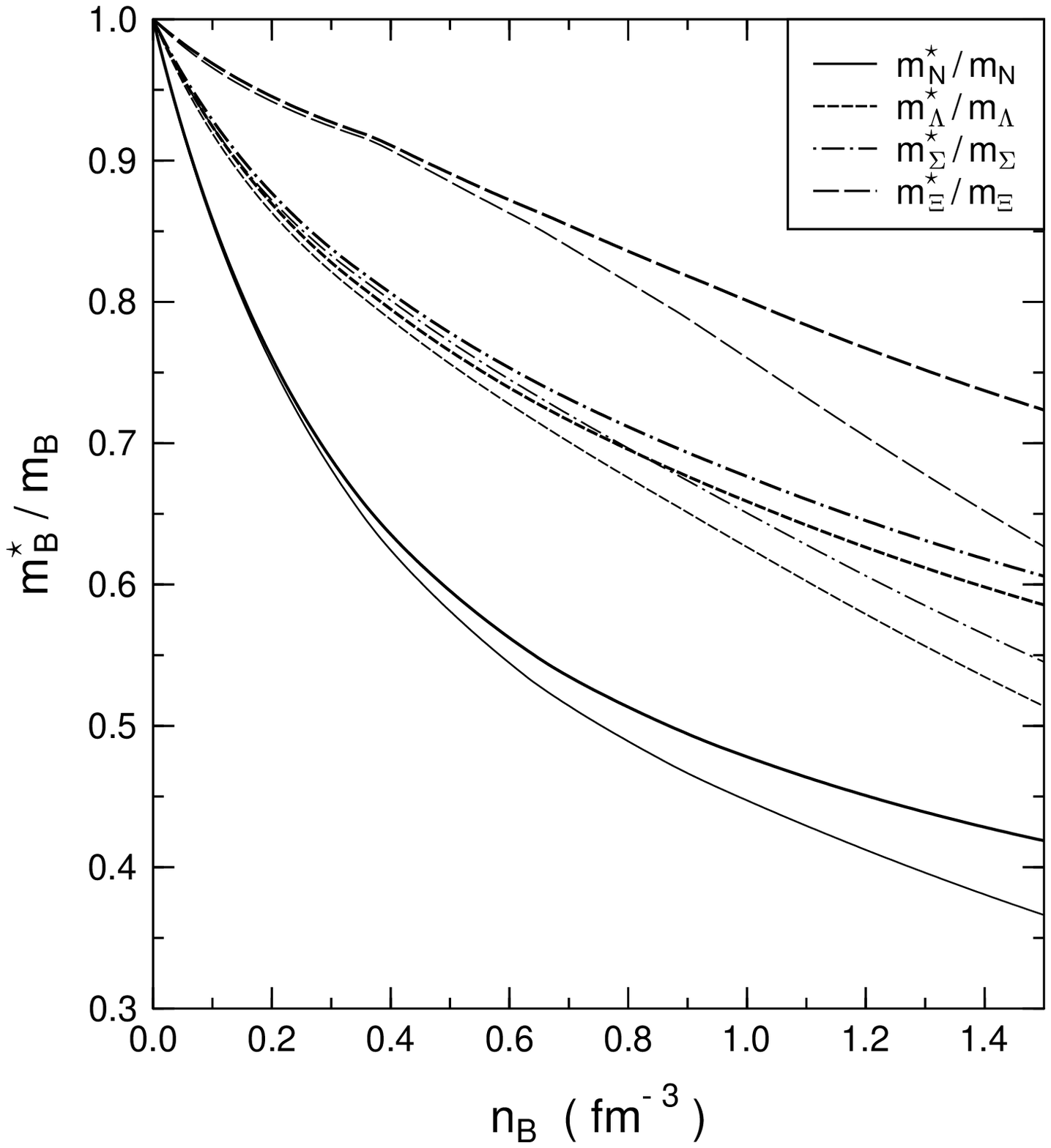}
}}

\vspace{-3.5cm}

\noindent{\small{
FIG. 4. The variation of effective  masses $m_B^*/m_B$ of the baryons
as a function of baryon density $n_B$ in the models QMCII (thick lines) 
and QHDII (thin lines).}}
\newpage
\vspace{-1cm}

{\centerline{
\epsfxsize=13cm
\epsfysize=15cm
\epsffile{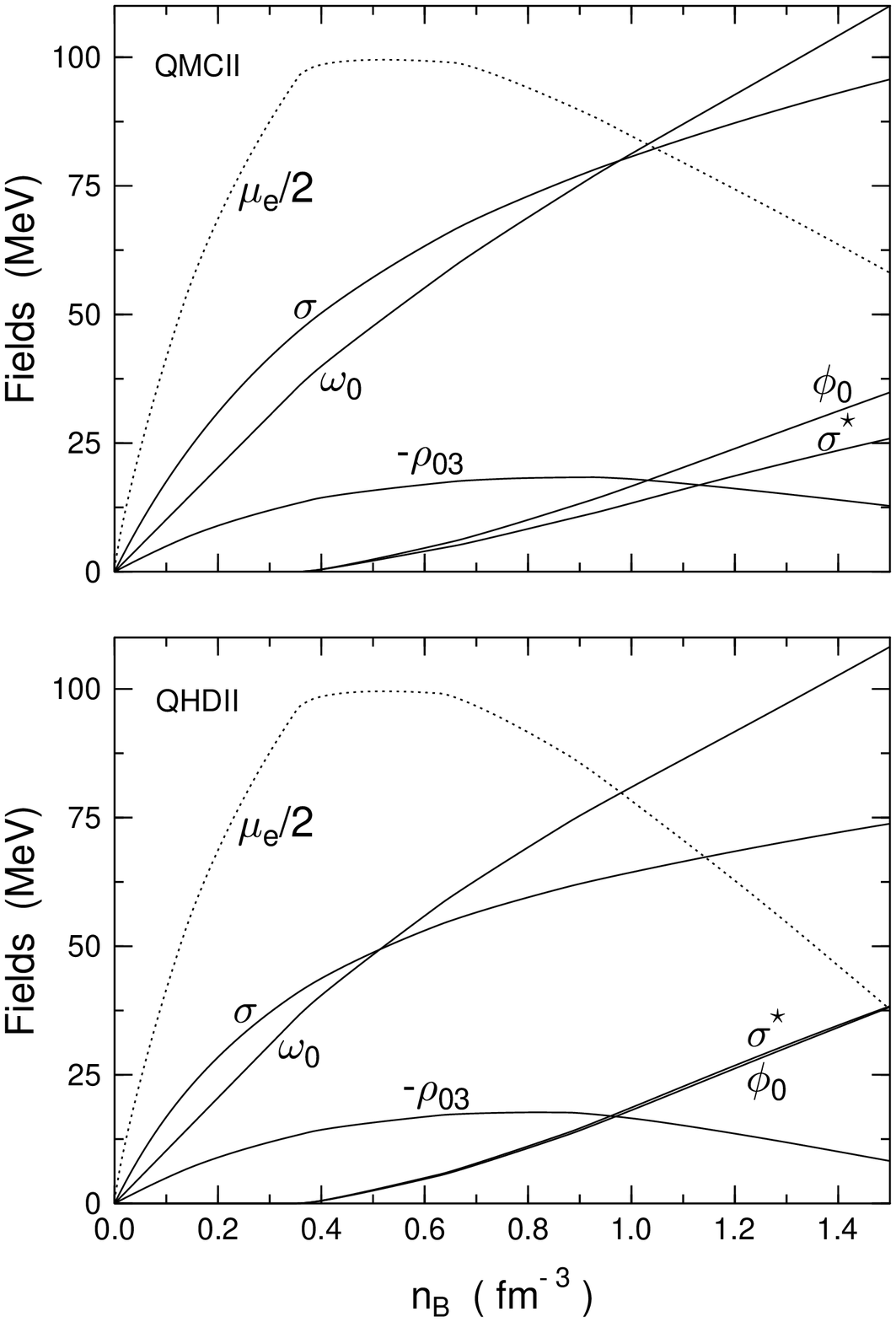}
}}

\vspace{-2cm}

\noindent{\small{
FIG. 5. The mean meson fields and the electrochemical potential versus the 
baryon density $n_B$ for the models QMCII (top panel) and QHDII (bottom panel).
}}
\newpage
\vspace{-1cm}

{\centerline{
\epsfxsize=13cm
\epsfysize=15cm
\epsffile{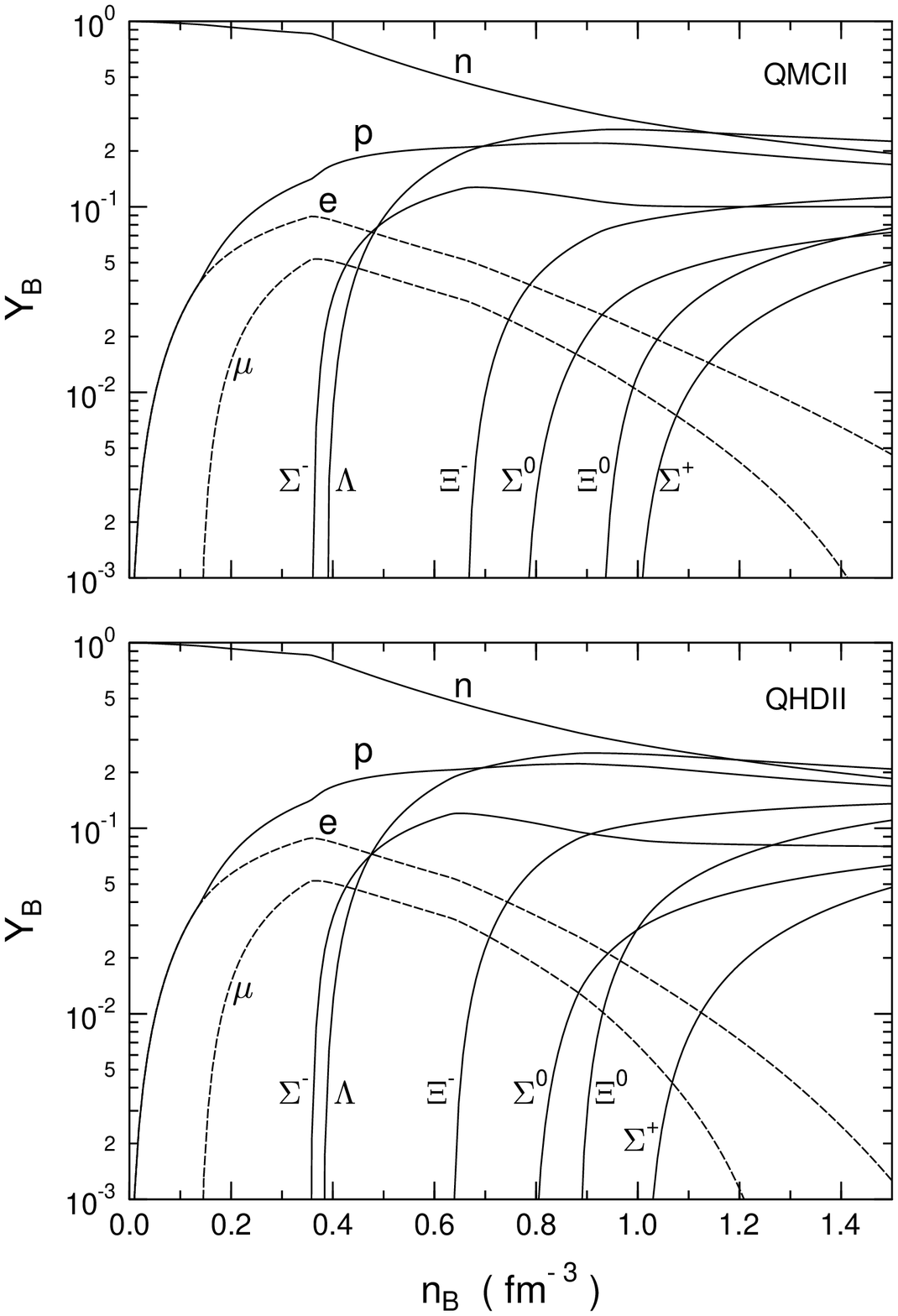}
}}

\vspace{-2cm}

\noindent{\small{
FIG. 6. The composition of neutron star matter with hyperons in the models 
QMCII (top panel) and QHDII (bottom panel).}}
\newpage
\vspace{-5cm}

{\centerline{
\epsfxsize=8cm
\epsfysize=9cm
\epsffile{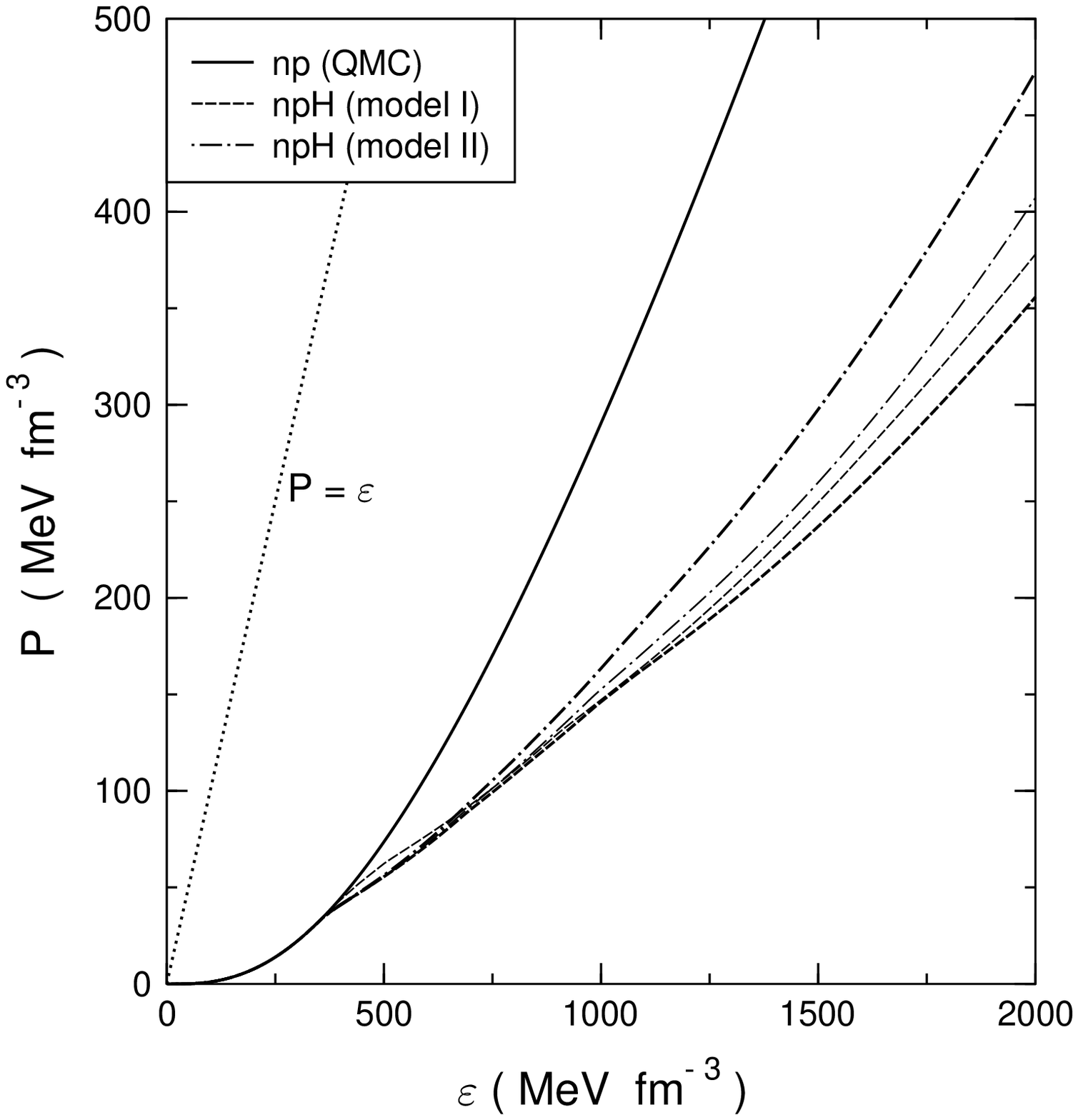}
}}

\vspace{-3.5cm}

\noindent{\small{
FIG. 7. The equation of state, pressure $P$ vs. energy density $\varepsilon$ 
for nucleons only (np) star (solid line) in the QMC model, and for stars with 
further inclusion of hyperons (npH) in models I (dashed lines) and in models II 
(dash-dotted) lines. The results are for QMC models (thick lines) and QHD 
models (thin lines). The causal limit ($P = \varepsilon$) is also shown.}}

{\centerline{
\epsfxsize=8cm
\epsfysize=8cm
\epsffile{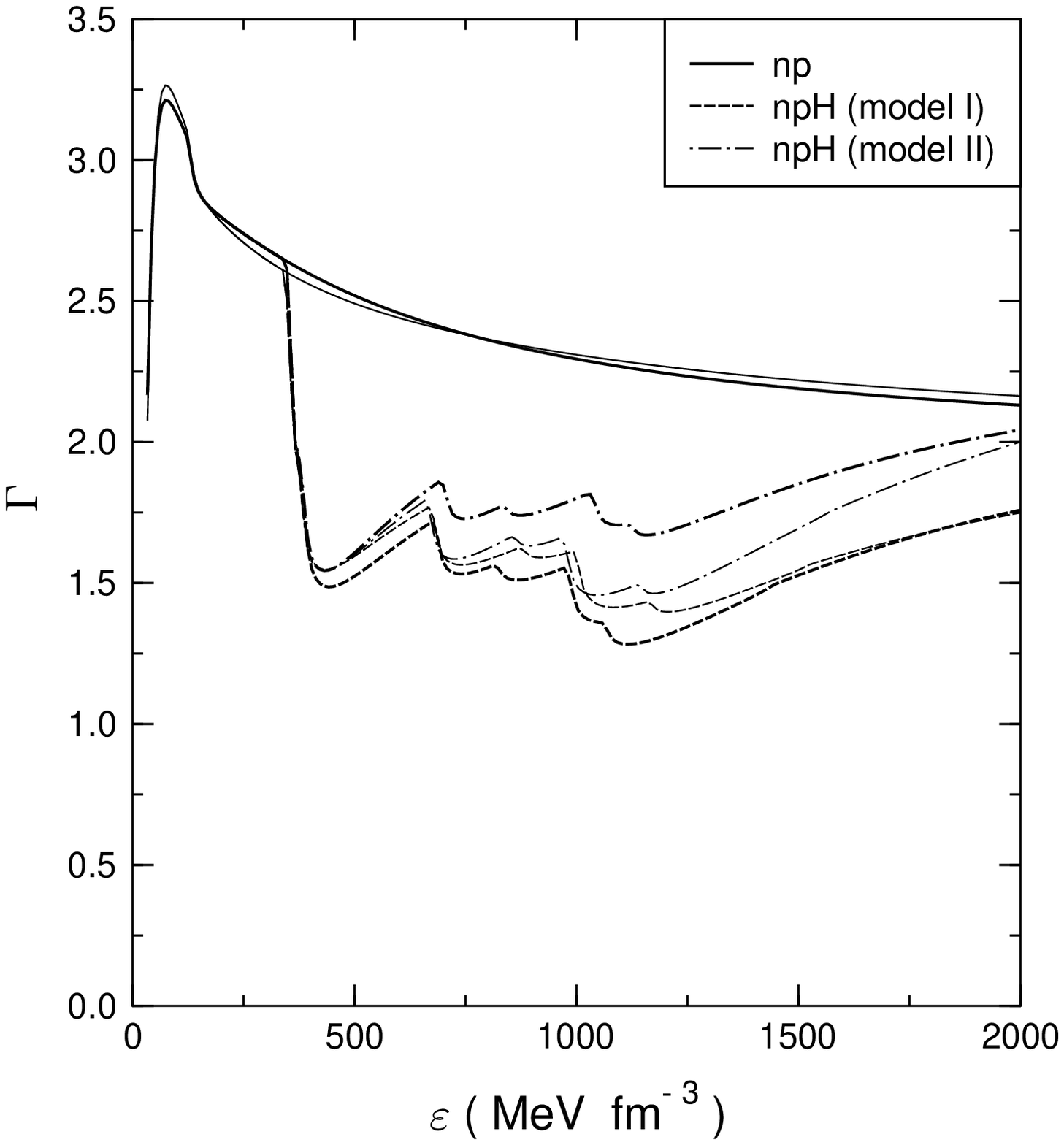}
}}

\vspace{-3.5cm}

\noindent{\small{
FIG. 8. The adiabatic index $\Gamma$ vs. the energy density $\varepsilon$ 
for nucleons only (np) star (solid lines), and for stars with further 
inclusion of hyperons (npH) in models I (dashed lines) and in models II 
(dash-dotted) lines. The results are for QMC models (thick lines) and QHD 
models (thin lines).}}
\newpage
\vspace{-5cm}

{\centerline{
\epsfxsize=8cm
\epsfysize=7cm
\epsffile{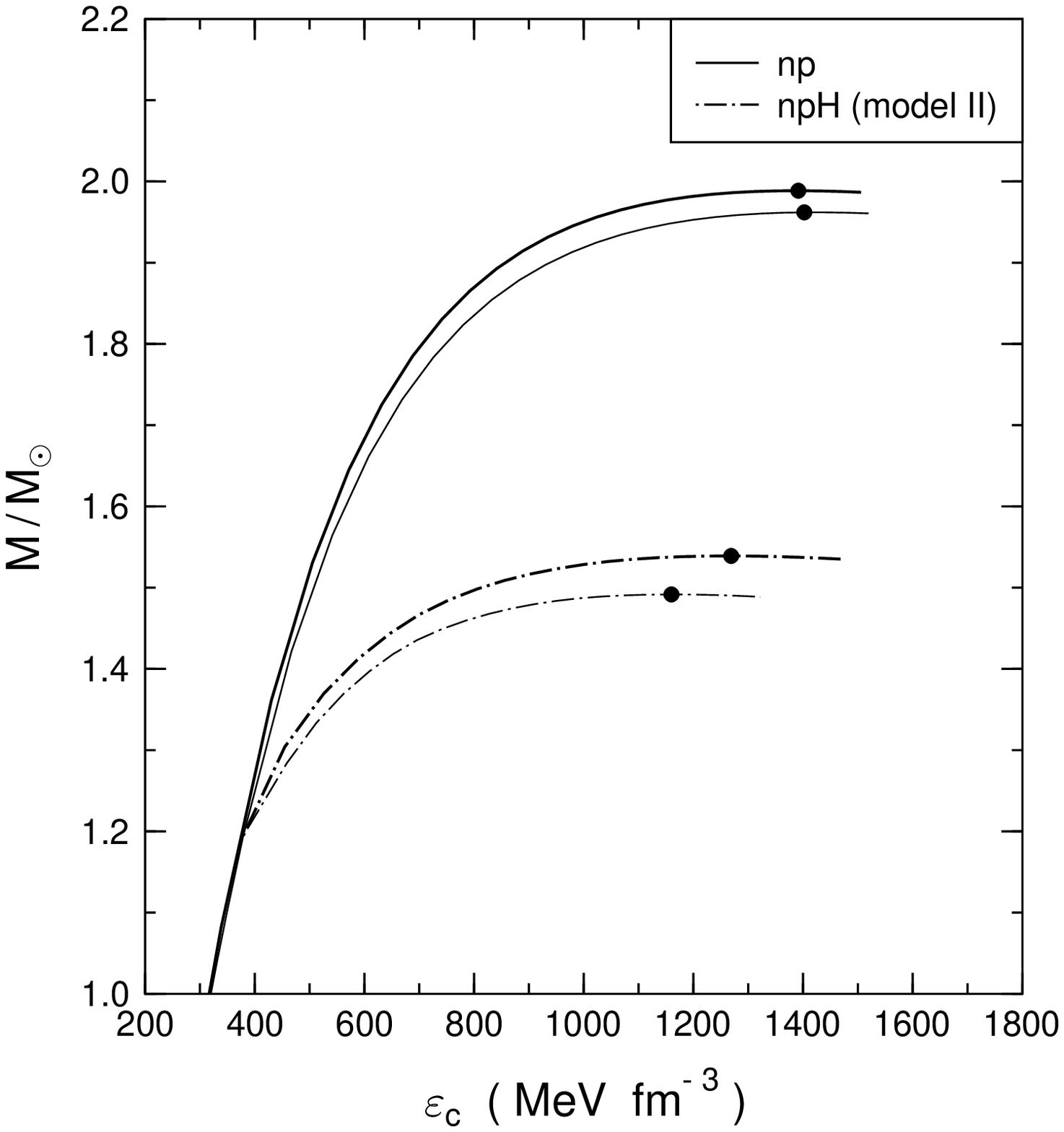}
}}

\vspace{-3.5cm}

\noindent{\small{
FIG. 9. The neutron star sequences near the limiting mass for nucleons only
(np) star (solid lines) and for stars with further inclusion of hyperons
(npH) in the model II (dash-dotted lines). The results are for the models
QMC (thick lines) and QHD (thin lines). The filled circles correspond to 
the maximum masses.}}

{\centerline{
\epsfxsize=8cm
\epsfysize=7cm
\epsffile{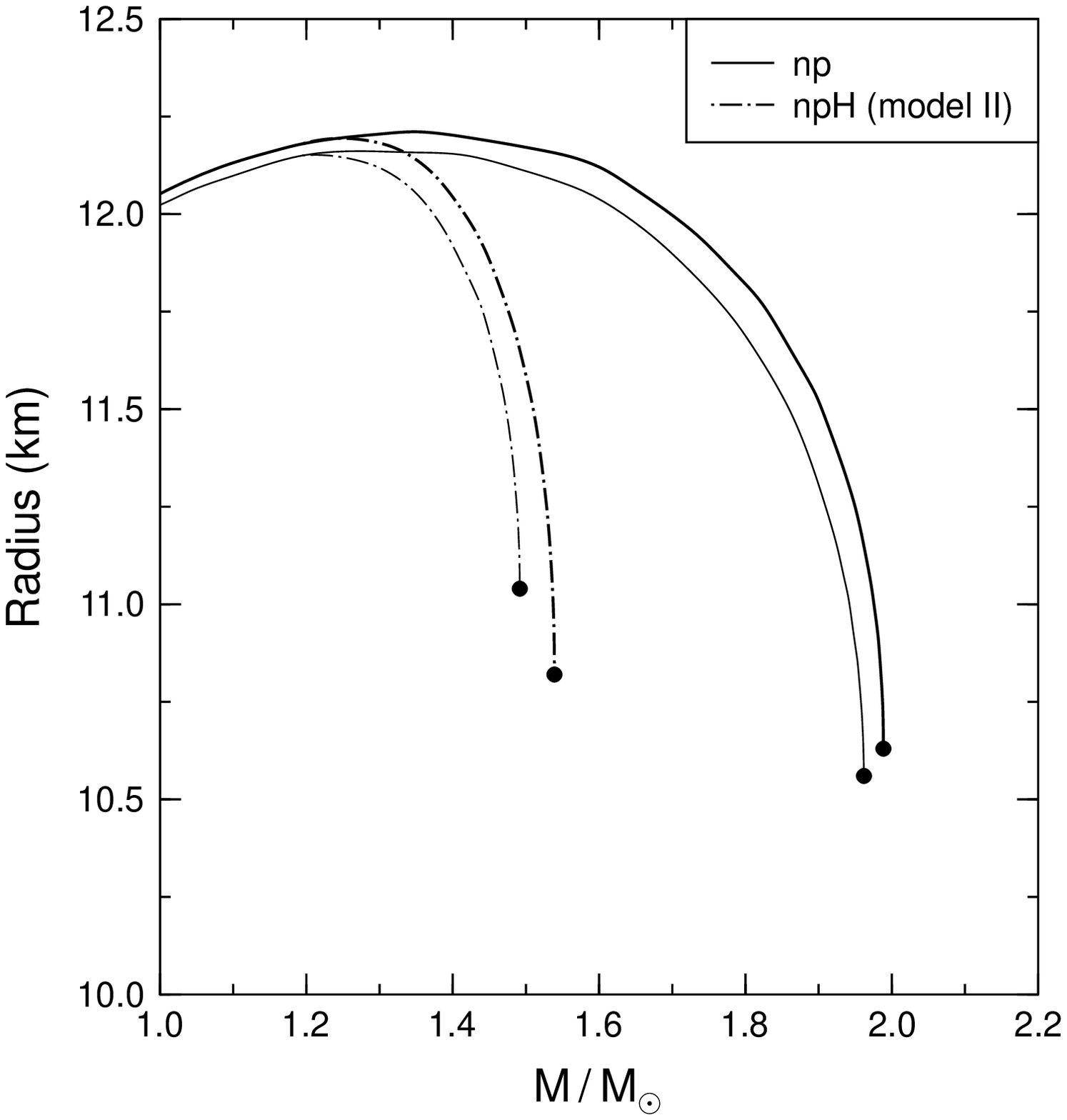}
}}

\vspace{-3.5cm}

\noindent{\small{
FIG. 10. The mass-radius relation for neutron star sequences near the 
limiting mass for nucleons only (np) star (solid lines) and for stars with 
further inclusion of hyperons (npH) in the model II (dash-dotted lines). 
The results are for the models QMC (thick lines) and QHD (thin lines). 
The filled circles represent the maximum masses.}}

\end{document}